\documentclass[aps,prb,twocolumn,citeautoscript,longbibliography,floatfix]{revtex4-2}

\usepackage{multirow}
\usepackage{graphicx}
\usepackage{longtable}
\usepackage[utf8]{inputenc}
\usepackage{epstopdf}
\usepackage{textcomp}
\usepackage[usenames,dvipsnames]{color}
\usepackage{amsbsy}
\usepackage{amsmath}
\usepackage{amsfonts}
\usepackage{calrsfs}
\usepackage{mathtools}
\usepackage{bbold}
\usepackage{braket}
\usepackage{array}
\usepackage{gensymb}
\usepackage{comment}

\DeclareMathAlphabet{\pazocal}{OMS}{zplm}{m}{n}
\newcommand{\Kb}{\pazocal{K}}
\newcommand{\Eb}{\pazocal{E}}
\newcommand{\Ab}{\pazocal{A}}
\newcommand{\Db}{\pazocal{D}}

\newcommand{\vc}[1]{\boldsymbol{\mathrm #1}}

\newcommand\varpm{\mathbin{\vcenter{\hbox{%
  \oalign{\hfil$\scriptstyle+$\hfil\cr
          \noalign{\kern-.3ex}
          $\scriptscriptstyle({-})$\cr}%
}}}}
\newcommand*\diff{\mathop{}\!\mathrm{d}}
\newcommand{\matr}[1]{\ensuremath{\underline{\mathbf{#1}}}}
\newcommand {\grbf}[1]{\mbox{\boldmath $ #1 $}}

 \allowdisplaybreaks

\begin{document}

\title{Atomically Sharp Magnetic Domain Walls beyond Micromagnetics}

\author{Guokai Liao}
\email[Corresponding author: ]{g.liao@fz-juelich.de, lgk@mail.ustc.edu. cn}
\affiliation{Peter Gr\"{u}nberg Institut, Forschungszentrum J\"{u}lich \& JARA, D-52425 J\"{u}lich, Germany}
\affiliation{International Center for Quantum Design of Functional Materials (ICQD), University of Science and Technology of China,
Hefei 230026, Anhui, China}
\author{Stefan Bl\"ugel}
\affiliation{Peter Gr\"{u}nberg Institut, Forschungszentrum J\"{u}lich \& JARA, D-52425 J\"{u}lich, Germany}
\affiliation{Institute for Theoretical Physics, RWTH Aachen University, D-52062 Aachen, Germany}

\begin{abstract}
Atomically narrow magnetic domain walls can arise in ultrathin magnetic films, van der Waals magnets, and systems with reduced effective exchange stiffness due to competing exchange interactions. As their width approaches the atomic lattice spacing, the continuum approximation underlying micromagnetics becomes increasingly unreliable. We investigate the structure and stability of such domain walls in a one-dimensional nearest-neighbor $J$--$K$--$D$ spin-lattice model incorporating Heisenberg exchange, uniaxial anisotropy, and Dzyaloshinskii–Moriya interaction (DMI). In the absence of DMI, we demonstrate analytically a continuous transition to an atomically sharp easy-axis wall at the critical exchange-to-anisotropy ratio $J/K=3/2$, characterized by the softening of a localized Hessian mode. 
Atomistic spin-dynamics simulations confirm this transition and reveal that fitting discrete wall profiles with the conventional micromagnetic solution, as commonly done in experimental analyses, underestimates the exchange interaction and overestimates the anisotropy. Finite DMI removes the perfectly collinear wall as a stationary configuration, replacing the transition by a smooth crossover toward a strongly localized but canted wall that approaches the collinear easy-axis configuration only asymptotically in the strong-anisotropy limit. We derive an analytical expression for the DMI-induced canting in this regime and demonstrate its strongly enhanced canting response near the critical ratio. These findings establish the limitations of micromagnetic descriptions of atomically narrow domain walls and suggest opportunities for controlling their structure through anisotropy and DMI.
\end{abstract}

\maketitle

\section{Introduction}
In a typical ferromagnet, the magnetization is distributed in ferromagnetic domains of different, typically  energy equivalent magnetization directions separated by magnetic domain walls. Although metastable, these domain-walls can have very long lifetimes and are of scientific and technological relevance. In the context of information technology, understanding the nature of domain walls in magnetic nanostructures has become an important issue. The field of spintronics~\cite{chappert:07} offers current induced controlled and quick transport~\cite{Boulle:11} of domain walls from input to output devices. This provides an interesting basis for building both low-power logic~\cite{Allwood:05.1,Xu:08,Luo:20} and memory~\cite{Parkin:04.1, Parkin:08.1,Ryu:13} domain-wall devices that can open a path to in-memory and beyond John von Neumann computing~\cite{Cui:20}.  
Domain walls (DW) in magnetic nanostructures and thin films arise from the interplay of the quantum-mechanical exchange interaction, described in the continuum limit by the spin stiffness, the magnetocrystalline anisotropy caused by the relativistic spin-orbit coupling (SOC), and the classical long-range magnetostatic contribution. The experimental characterization of the domain-wall profile provides information about these quantities~\cite{Pratzer:01.1,Pratzer:03.1,Yang:22}.  In ultrathin films and multilayers, surface and interface contributions to the magnetic anisotropy often dominate over bulk contributions at film thicknesses below approximately 2~nm, frequently favoring an out-of-plane magnetic easy axis. Although the magnetostatic dipole-dipole interactions are intrinsically non-local and long-ranged their contribution in ultrathin films can often be approximated by a local effective anisotropy term~\cite{Heide:08.1}.

High information-density requirements of domain-wall devices~\cite{chappert:07} drive research in the direction of narrow domain walls ideally below a width of 10~nm, which can be typically achieved in bulk magnets with reduced effective exchange stiffness due to competing exchange interactions~\cite{Allen:26} or in ultrathin films at the limit of a few  atomic layers of magnetic materials. Then, the ferromagnetic volume, thus the number of magnetic neighbors and subsequently the exchange stiffness $\Ab$ decreases. Under the same circumstances, the surface contribution to the magnetic anisotropy dominates over the volume one  and the magnetic anisotropy $\Kb$ increases leading to domain-wall widths $w\propto \sqrt{\Ab/\Kb}$ of atomistic or nanometer scale~\cite{Nakamura:05,Heide:08.1}. In the past, these ultrathin films and the small domain walls have been achieved by the deposition of magnetic materials on heavy-metal substrates, \textit{e.g.}\ the observed narrow domain walls in Fe films on W(110)~\cite{Pratzer:01.1,Pratzer:03.1}. In this context, the discovery of two-dimensional (2D) van der Waals magnets~\cite{Huang:17,Gong:17,Guguchia:18,Klein:18} is a game changer as it has opened up a completely new frontier of ultrathin magnetic films with sub-10-nm magnetic domain walls~\cite{Purbawati:20,Pei:21,Abdul-Wahab:21,Alliati:22,Yang:22}. 

In general, the rotation of the magnetization across a domain wall has no preferred handedness, and walls with clockwise and counterclockwise rotational sense are energetically equivalent. In ultrathin films, multilayers, and heterostructures, however, broken inversion symmetry together with spin--orbit coupling gives rise to the Dzyaloshinskii--Moriya interaction (DMI), characterized by the strength $\Db$~\cite{Heide:08.1}. Provided that the DMI remains below the critical threshold $D_\mathrm{c}=2 \sqrt{\Ab\Kb}$ for a modulated magnetic ground state, the ferromagnetic domains remain stable but DMI lifts the degeneracy between the two possible rotational senses of the domain wall texture. Besides selecting the chirality of the wall magnetization, the DMI selects the wall type, \textit{e.g.}\ a N\'eel~\cite{Neel:1955} rather than a Bloch wall~\cite{Bloch:1932}. Such chiral domain walls have been observed experimentally in a variety of ultrathin magnetic systems \cite{Kubetzka:03,Chen:13,Chen:17,Birch:22}. Besides determining the type and rotational sense of the wall, the DMI strongly affects its dynamics: chiral domain walls can strongly increase the threshold for the conventional Walker breakdown and thereby sustain substantially higher current-driven velocities \cite{Thiaville:12,Ryu:13}.

The conventional description of the domain-wall profile and its width $w$ is based on micromagnetics, which assumes that the magnetization varies smoothly on the scale of the atomic lattice. As domain walls become increasingly narrow, however, this fundamental assumption eventually breaks down. The experimentally observed reduction of domain-wall widths toward only a few lattice constants, together with the occurrence of DMI-induced chiral walls in ultrathin magnetic systems, therefore motivates a description in terms of discrete atomic spins.

In this work, we investigate how the competition between exchange and magnetic anisotropy determines the structure and stability of domain walls at the atomic scale, and how this behavior is modified by DMI. We find that the discrete exchange–anisotropy ($J$–$K$) model exhibits a continuous transition at $J/K=3/2$ to an atomically sharp easy-axis wall (EAS), in which all spins are aligned parallel or antiparallel to the easy axis and the magnetization reverses abruptly between two neighboring lattice sites. The transition is characterized by the vanishing of the  lowest  eigenvalue of the energy Hessian. Near the critical point, fitting the magnetic domain-wall profile with the continuum model leads  to underestimation of the exchange interaction and overestimation of the magnetic anisotropy. A finite Dzyaloshinskii--Moriya interaction (DMI), however, removes the stationary easy-axis state, lifts the soft mode, and replaces the transition by a smooth crossover.  
The paper is organized as follows. First, we introduce the conventional micromagnetic and discrete spin-lattice models and derive the critical exchange-to-anisotropy ratio at which an infinitely long chain undergoes a continuous transition from a canted domain wall to an atomically sharp easy-axis wall. We then compare numerically relaxed configurations of a 100-spin chain, obtained with Spirit~\cite{spirit,mueller_spirit_2019} with the continuum domain-wall profile. This comparison yields apparent continuum parameters and exposes the energetic and profile errors of the continuum approximation near the atomically sharp-wall transition. Finally, we investigate the effects of finite DMI on the structure and stability of the domain wall through an analysis of the Hessian spectrum, the minimum out-of-plane magnetization, and the linear response of the wall in the strong-anisotropy regime.

\section{Models}
We investigate the structure, energetics, and stability of narrow magnetic domain walls at zero temperature using two complementary descriptions: the micromagnetic continuum model and a discrete atomistic spin-lattice model incorporating exchange, uniaxial anisotropy, and Dzyaloshinskii–Moriya interaction.

\subsection{Continuum Model}
\label{sec:continuum}

We consider two semi-infinite magnetic domains
with opposite out-of-plane magnetizations, separated by a straight
domain wall centered at $x=0$. Assuming translational invariance
along the wall, the system can be described by the well-established
one-dimensional (1D) micromagnetic energy functional
\cite{Hubert:98.1},
\begin{equation}
\Eb[\vc{m}]
=\int \diff x\,
\left[
\Ab\left(\frac{\diff\vc{m}}{\diff x}\right)^2
-\Kb m_z^2
\right]\, ,
\label{eq:micromagnetic_Em}
\end{equation}
where $\vc{m}(x)$ is the normalized magnetization,
$|\vc{m}(x)|=1$. The first term describes the exchange
interaction in the continuum approximation, characterized by
the exchange stiffness $\Ab>0$, whereas the second term
represents the uniaxial magnetic anisotropy $\Kb>0$,
which favors magnetization along the out-of-plane $z$ axis.
The magnetostatic contribution is assumed to be included
in the effective anisotropy.

In the absence of interactions that distinguish different
rotation planes, the magnetization can be confined to a
fixed plane without loss of generality. The two conventional
domain-wall configurations are the N\'eel wall, for which
the magnetization rotates in the $xz$ plane,
$\vc{m}_{\mathrm N}(x)=\bigl(\cos\varphi(x),0,\sin\varphi(x)\bigr)$,
and the Bloch wall, for which the rotation takes place
in the $yz$ plane, 
$\vc{m}_{\mathrm B}(x) =\bigl(0,\cos\varphi(x),\sin\varphi(x)\bigr)$.
Within the exchange--anisotropy model of
\eqref{eq:micromagnetic_Em}, these configurations,
as well as walls with any intermediate fixed rotation
plane, are energetically degenerate. The angle
$\varphi(x)$ describes the local magnetization direction
relative to the corresponding in-plane axis.

For either wall type, \eqref{eq:micromagnetic_Em}
reduces to
\begin{equation}
\Eb[\varphi]
=\int \diff x\,
\left[
\Ab\left(\frac{\diff\varphi}{\diff x}\right)^2
-\Kb\sin^2\varphi
\right].
\label{eq:micromagnetic_Ephi}
\end{equation}
The equilibrium domain-wall profile are the well-known analytical solutions  \cite{Landau:35.1} 
\begin{eqnarray}
\sin\varphi(x)&=&\tanh\left(\frac{x}{\xi}\right),\nonumber \\
 \text{and}\qquad\qquad &&  \label{eq:micromagnetic_DW-profile} \\
\cos\varphi(x)&=&
\operatorname{sech}\left(\frac{x}{\xi}\right)=\pm \sqrt{1-\tanh^2\left(\frac{x}{\xi}\right)} \nonumber \, ,
\end{eqnarray}
of  the corresponding Euler--Lagrange equations that minimizes this
functional subject to the boundary conditions
$m_z=-1$ for $x\rightarrow -\infty$ and $m_z=1$ for $x\rightarrow\infty$
corresponding to 
\begin{equation}
\varphi(-\infty)=-\frac{\pi}{2},
\qquad
\varphi(+\infty)=\frac{\pi}{2}, ,
\end{equation}
with $\varphi(x)\in[-\pi/2,\pi/2]$ being the angle between the magnetization direction and the $x$ direction. The two possible signs of the transverse magnetization component correspond to opposite senses of rotation. $\xi=\sqrt{{\Ab}/{\Kb}}$ is the characteristic domain-wall length and for the domain-wall width, defined as $w=2\xi$, we obtain
\begin{equation}
w=2\xi= 2\sqrt{\frac{\Ab}{\Kb}}.
\label{eq:micromagnetic_width}
\end{equation}
The corresponding domain-wall energy, measured relative
to the uniformly magnetized domains, is
\begin{equation}
E_{\mathrm{DW}}^{(0)}
=4\sqrt{\Ab\Kb}.
\label{eq:micromagnetic_DW_energy}
\end{equation}
Thus, within the continuum model, the domain-wall
width decreases continuously with increasing anisotropy
but remains finite for any finite $\Ab/\Kb$.

The degeneracy between Bloch and N\'eel walls can be
lifted by the Dzyaloshinskii--Moriya interaction (DMI).
For an interfacial DMI with the symmetry considered
here, its contribution to the one-dimensional
micromagnetic energy is
\begin{equation}
\Eb_{\mathrm{DMI}}[\vc{m}]
=\Db\int\diff x\,
\left(
m_z\frac{\diff m_x}{\diff x}
-m_x\frac{\diff m_z}{\diff x}
\right),
\label{eq:micromagnetic_DMI}
\end{equation}
where $\Db$ denotes the micromagnetic DMI coefficient.
This contribution vanishes for a Bloch wall, whereas
for a N\'eel wall it reduces to
\begin{equation}
\Eb_{\mathrm{DMI}}[\varphi]
=-\Db\int\diff x\,
\frac{\diff\varphi}{\diff x}
=-\Db\bigl[\varphi(+\infty)-\varphi(-\infty)\bigr].
\label{eq:micromagnetic_DMI_phi}
\end{equation}
Since the DMI contribution is a total derivative,
it does not modify the Euler--Lagrange equation
for the equilibrium profile under fixed boundary
conditions. Consequently, the N\'eel-wall profile
and its characteristic width remain unchanged,
whereas the domain-wall energy becomes
\begin{equation}
E_{\mathrm{DW}}^{\mathrm N}
=4\sqrt{\Ab\Kb}\pm\pi\Db.
\label{eq:micromagnetic_DW_DMI_energy}
\end{equation}
The two signs correspond to opposite wall chiralities,
with the energetically preferred chirality selected
by the sign of $\Db$. Thus, within the continuum
model, DMI selects the rotational sense of a N\'eel
wall without changing its equilibrium profile.

\subsection{Discrete Spin-Lattice Model}

We consider an atomistic nearest-neighbor spin-lattice model
\begin{equation}
E[\vc{S}]=-J\sum_{\langle ij\rangle}\vc{S}_i\cdot\vc{S}_{j}
-D_y\sum_{\langle ij\rangle}\hat{\vc y}\cdot(\vc{S}_i\times\vc{S}_{j})
-K\sum_i\vc{S}_{i|z}^2
\label{eq:spin-model_E}
\end{equation}
of a 1D chain of atoms separated by a lattice constant $a$ with local classical spins  $\vc{S}_i$ of unity length, $\vert \vc{S}_i\vert=1$ $ \forall i$, placed on sites $i$. The first term describes the Heisenberg exchange interaction between  nearest-neighbor pairs of atoms, indicated by $\langle ij\rangle$, designating bonds between neighboring lattice sites, counting each bond only once,  with an exchange parameter $J$ describing the ferromagnetic state, $J\ge 0$. The second term is the Dzyaloshinskii--Moriya interaction (DMI) with $\vc D=D_y\hat{\vc y}$; its sign fixes the chirality of the planar wall (for $D_y>0$, the handedness of a spiral with positive vector chirality $c_y=(\mathbf S_i\times\mathbf S_{i+1})_y>0$ has lower energy). The third term describes the on-site uniaxial out-of-plane magnetic anisotropy with parameter $K>0$. $E[\vc{S}]$ is the energy of the 1D chain of atoms, while $\Eb[\vc{m}]$ is the energy per materials cross section $F$ normal to $x$ direction, and both are related as $E[\vc{S}]= F \Eb[\vc{m}]$.  

In the long wave-length limit, \textit{i.e.}\ considering that the magnetization of the system varies slowly in space, we relate the discrete spin $\vc{S}_j$ value at site $j$ to the continuous function $\vc{m}(x)$ as $\vc{S}_j=\vc{m}(ja)$ and obtain after Taylor expansion $\vc{S}_j \simeq \vc{m}(0) + ja \,\text{d}\vc{m}/\text{d}x|_{(0)} +1/2 j^2a^2 \,\text{d}^2\vc{m}/\text{d}x^2|_{(0)}$ the following relations between the parameters in the micromagnetic and spin-lattice model 
\begin{equation}
\Ab=\frac{a}{2F} J \quad\text{,}\quad\Db=\frac{1}{F}D_y\quad\text{and}\quad  \Kb = \frac{1}{aF} K\, .
\label{eq:rel_micro-to-atomistic_parameters} 
\end{equation}

\begin{figure}[t]
\centering
\includegraphics[width=0.95\columnwidth]{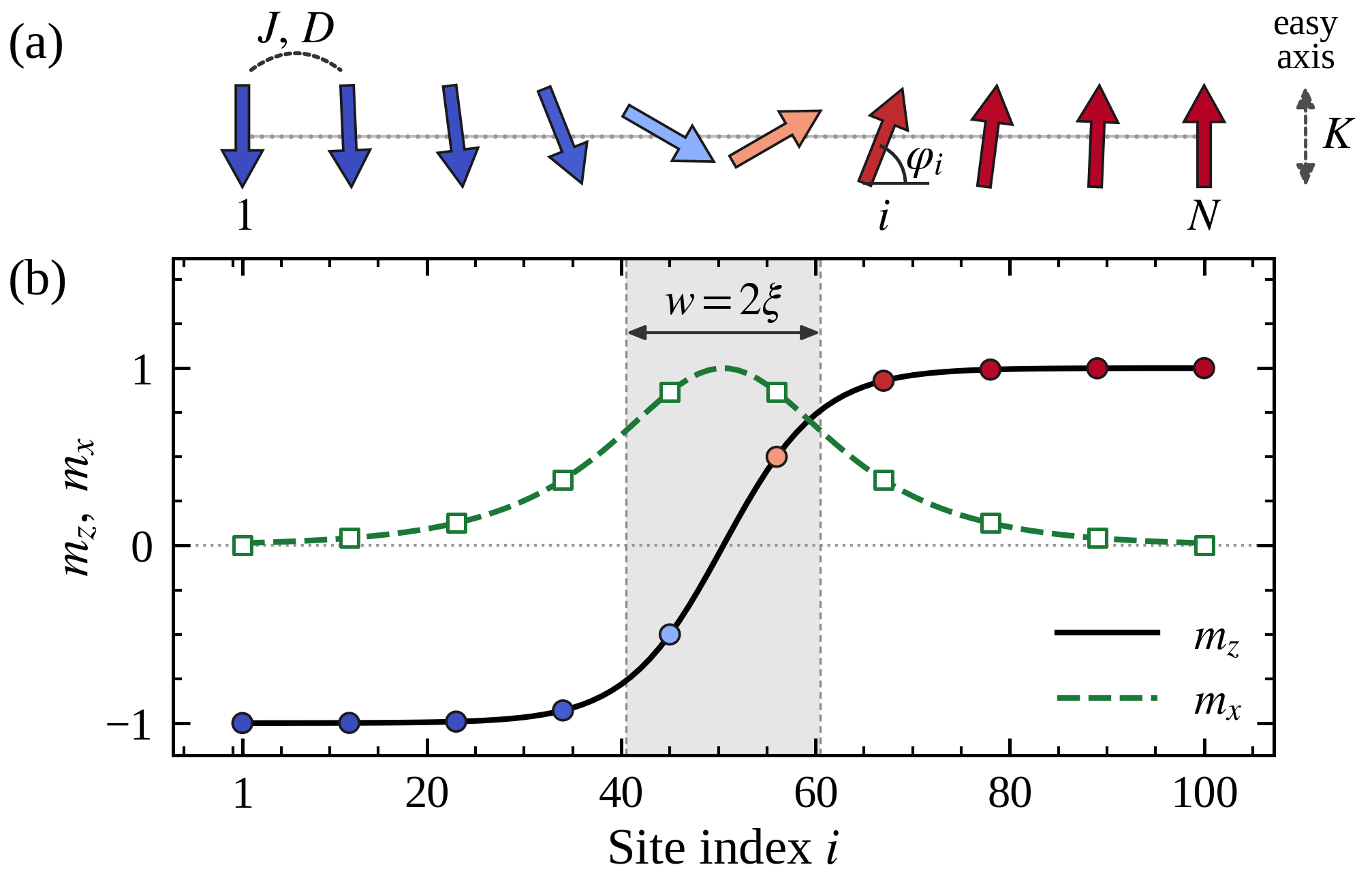}
\caption{Graphical representations of one-dimensional continuum and discrete domain walls. The upper part presents the discrete model with local spins of unit length at lattice sites $i$, where the orientation of spins is characterized by an angle $\varphi_i$, which increases with increasing site index, representing a counterclockwise rotating wall in the $xz$ plane.  The typical continuum profiles for magnetizations $m_x$ and $m_z$ are shown below.}
\label{fig:1D-wall}
\end{figure}

In analogy to Eq.~\eqref{eq:micromagnetic_Ephi} we write the spin vector $\vc{S}_i= (\cos\varphi_i, 0 , \sin\varphi_i)$ of unit length  in terms of the angles $\varphi_i$ at sites $i$ with respect to the chain direction and can thus rewrite the energy of the atomistic spin-model  \eqref{eq:spin-model_E} as 
\begin{equation}
E[\varphi]=-J\sum_{\langle ij\rangle}^N\cos\Delta_{ij}
+D_y\sum_{\langle ij\rangle}^N\sin\Delta_{ij}
-K\sum_i^N\sin^2\varphi_i\, ,
\label{eq:spin-model_Ephi}
\end{equation}
where $\Delta_{ij}=\varphi_j-\varphi_i$ and  $N$ is the number of spins. The domain-wall problem is consistent with the twisted boundary conditions in the system,  which are expressed by 
\begin{equation}
\varphi_1=-\frac{\pi}{2}\quad\text{and}\quad \varphi_N=\frac{\pi}{2} \, . 
\label{eq:spin-model_BC}
\end{equation} 
Thus, the first spin is fixed in the downward direction and the last spin in the upward direction. Figure~\ref{fig:1D-wall} gives a graphical representation of the one-dimensional continuum and discrete domain walls. The upper part presents the discrete model, where the orientation of spins is characterized by an angle $\varphi_i$, whereas a typical continuum profile is shown below. 

With the parametrization $\mathbf S_i=(\cos\varphi_i,0,\sin\varphi_i)$, the $y$ component of the nearest-neighbor vector chirality is
\begin{equation}
c_{i,y}=(\mathbf S_i\times\mathbf S_{i+1})_y
=-\sin\Delta_i,
\qquad \Delta_i=\varphi_{i+1}-\varphi_i.
\end{equation}
Accordingly, the DMI term in the Hamiltonian \eqref{eq:spin-model_E}, favors positive vector chirality for $D_y>0$, corresponding to negative angular increments $\Delta_i$ for $|\Delta_i|<\pi$. In the $xz$-plane representation of Fig.~\ref{fig:1D-wall}, with $+\hat{\mathbf x}$ pointing right and $+\hat{\mathbf z}$ upward, this corresponds to a clockwise rotation of the spins along the chain.

\section{Easy-axis transition in the $J$--$K$ model}
\label{sec:J-K-Model}

We now investigate how the interaction parameters and system size influence the domain wall structure. We first present analytical results and then compare them with numerical calculations. Throughout this section we set $D_y=0$ in Eqs.~\eqref{eq:spin-model_E} and~\eqref{eq:spin-model_Ephi}.

\subsection{Analytical Results}

We first examine the stability of atomically sharp easy-axis domain walls in the discrete spin-lattice model. In the continuum description, the domain-wall width decreases continuously with increasing magnetic anisotropy but remains finite for any finite ratio of exchange stiffness to anisotropy. In a discrete spin chain, however, the rotation of the magnetization can become confined to a single bond, giving rise to an atomically sharp easy-axis state (EAS), in which all spins are aligned parallel or antiparallel to the easy axis. The stability of this state is governed by the competition between the ferromagnetic exchange interaction $J$, which favors a gradual rotation of neighboring spins, and the uniaxial anisotropy $K$, which favors collinear alignment along the easy axis.

To illustrate this competition, we first consider a chain of four spins with fixed antiparallel boundary spins (for details see \ref{A_ssec:4-states}). The two inner spins are allowed to rotate, and their equilibrium orientations follow from minimizing the total energy \eqref{eq:spin-model_Ephi}. Besides the domain-wall state (DWS), characterized by finite canting angles, the energy admits an EAS in which the magnetization reverses abruptly between the two central spins. The analytical solution reveals a critical exchange-to-anisotropy ratio
$ \alpha_\mathrm{c}(4)=2(\xi_\mathrm{c}(4)/a)^2=\left({J}/{K}\right)_\mathrm{c}=2$.
For \(\alpha<\alpha_\mathrm{c}(4)\), the EAS is stable, whereas for \(\alpha>\alpha_\mathrm{c}(4)\), the energy is minimized by a canted DWS. At the critical ratio, the canting angles vanish continuously, indicating a continuous transition between the two configurations.

Increasing the chain length to six spins introduces additional degrees of freedom and modifies the critical ratio. The stability analysis yields $\alpha_\mathrm{c}(6)\simeq1.535$, demonstrating that the transition persists but that its position in the $(J, K)$ parameter space depends on the number of spins. This raises the question of whether the critical ratio approaches a finite value in the thermodynamic limit or whether the atomically sharp state is merely a finite-size effect.

To address this question, we examine the stability of the EAS for a chain of arbitrary even length \(N\). Expanding the energy to second order in small angular deviations \(\delta\varphi_i\) from the collinear configuration gives
\begin{equation}
E=E_{\mathrm{EAS}}+
\frac12\sum_{ij}H_{ij}\delta\varphi_i\delta\varphi_j
+\mathcal O(\delta\varphi^3),
\end{equation}
where
\begin{equation}
H_{ij}=
\left.
\frac{\partial^2 E}
{\partial\varphi_i\partial\varphi_j}
\right|_{\mathrm{EAS}}
\end{equation}
is the matrix elements of the Hessian matrix $\matr{H}$. The EAS is locally stable when all eigenvalues of $\matr{H}$ are positive. The transition occurs when its lowest eigenvalue vanishes, signaling the onset of an instability toward a noncollinear domain-wall configuration.

For nearest-neighbor exchange, the Hessian has a tridiagonal structure, \eqref{eq:app_DW-Hessian}, with a modification of the matrix elements associated with the antiparallel spins at the wall center. This matrix structure allows the stability condition to be evaluated analytically using Chebyshev polynomials of the second kind \eqref{eq:app_recurrence}, \eqref{eq:Cheby2nd_representation}. The detailed derivation is presented in Appendix A. In the thermodynamic limit, we obtain
\begin{equation}
\alpha_\mathrm{c}(\infty)=\frac{2}{a^2}\xi^2_\mathrm{c}(\infty)=\left(\frac{J}{K}\right)_\mathrm{c}=\frac32.
\label{eq:critical_alpha_infinity}
\end{equation}

Thus, an atomically sharp EAS remains stable even in an infinitely long spin chain, provided that $J/K<3/2$. The instability is driven by a mode localized at the domain wall, whose Hessian eigenvalue, $\varepsilon_{\mathrm{loc}}=2K-4J/3$, lies below the spectrum of the uniform ferromagnetic chain and vanishes at the critical ratio. For $J/K>3/2$, this eigenvalue becomes negative, and the EAS becomes unstable toward finite canting of the spins near the wall center. The small difference between $\alpha_\mathrm{c}(6)$ and $\alpha_\mathrm{c}(\infty)$ demonstrates the rapid convergence of the critical ratio with increasing system size.

The existence of this critical ratio highlights a fundamental difference between the discrete and continuum descriptions. In the micromagnetic model, the exchange energy increases without bound as the domain-wall width approaches zero, preventing the formation of an exactly sharp wall. In the discrete model, by contrast, the exchange energy associated with an antiparallel pair of neighboring spins remains finite. Consequently, sufficiently strong anisotropy can stabilize a magnetization reversal across a single atomic bond. The resulting transition from a canted DWS to an atomically sharp EAS is therefore an intrinsic lattice effect that cannot be captured by the conventional continuum approximation.

\subsection{Numerical Results}
Using Landau--Lifshitz--Gilbert-type spin-dynamics algorithms at $T=0$ as implemented in the Spirit code~\cite{spirit,mueller_spirit_2019}, we obtain numerical results for a chain of 100 spins, which can be compared with and fitted to the magnetization profile of the conventional continuum theory. The first and last spins are fixed by Eq.~\eqref{eq:spin-model_BC}, and the remaining spins are relaxed until the largest absolute component of the residual torque is below $10^{-13}$ in the energy units used by Spirit. The critical value $\alpha_\mathrm{c}$ can be determined accurately within $0.1\%$. Finite-size effects only have to be taken into account for system sizes of 10 sites or less, demonstrating rapid convergence.

\begin{figure}[t]
\centering
\includegraphics[width=0.83\columnwidth]{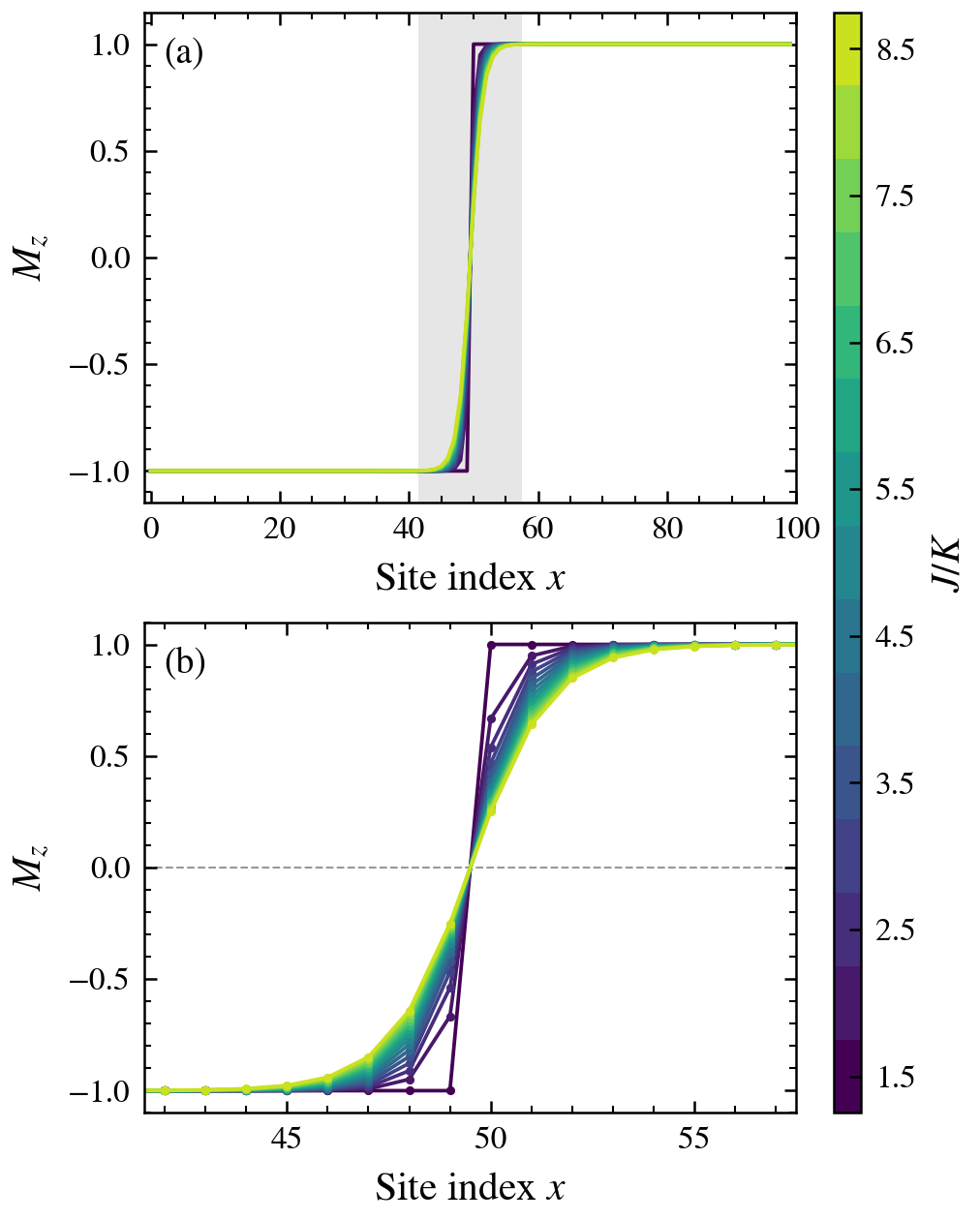}
\caption{Relaxed domain-wall profiles for a chain of 100 spins at $D_y=0$. Panel (a) shows $M_z$ for several ratios $J/K$, and panel (b) enlarges the wall center. The wall narrows as $J/K$ decreases and becomes atomically sharp at $J/K=3/2$.}
\label{fig:profiles}
\end{figure}

In order to compare the continuum theory with the lattice model, we simulate a discrete system and generate an equilibrium domain wall using the Hamiltonian in Eq.~\eqref{eq:spin-model_E}. Figure~\ref{fig:profiles} shows the resulting profiles for several values of $\alpha=J/K$. For large $\alpha$, the domain wall is broad. As $\alpha$ decreases, the rotation becomes localized to fewer sites and reaches the easy-axis state at $\alpha=3/2$, in agreement with Eq.~\eqref{eq:critical_alpha_infinity}.

The shape of the discrete domain wall can then be fitted to the profile predicted by continuum theory in Eq.~\eqref{eq:micromagnetic_DW-profile}. The energy and wall width of the best fit determine the corresponding $\Ab$ and $\Kb$ and, through Eq.~\eqref{eq:rel_micro-to-atomistic_parameters}, the explicit lattice model parameters $J_\mathrm{CT}$ and $K_\mathrm{CT}$. These apparent values can then be compared with the actual $J$ and $K$ used to generate the domain wall in the discrete model.

\begin{figure}[t]
\centering
\includegraphics[width=0.92\columnwidth]{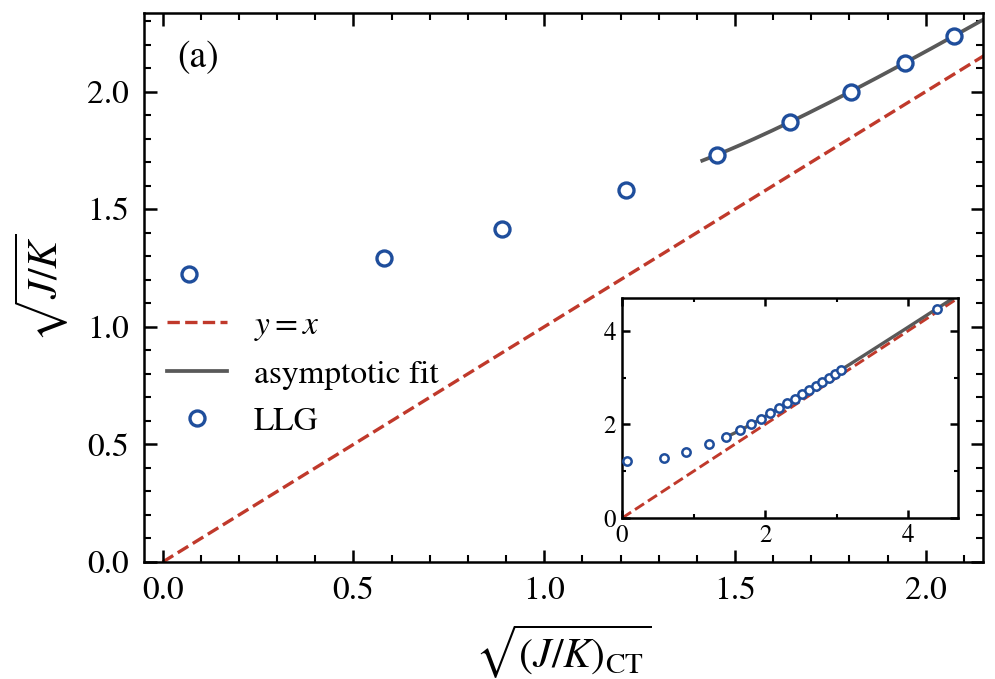}
\caption{Square root of the microscopic ratio $J/K$ used in the spin-dynamics simulations as a function of the corresponding value obtained by fitting Eq.~\eqref{eq:micromagnetic_DW-profile}. The dashed line marks equality and the solid line is the asymptotic fit. The inset extends the comparison to broader walls. The absolute discrepancy of $\sqrt{J/K}$ vanishes in the continuum limit.}
\label{fig:ratiofit}
\end{figure}

These results are presented in Fig.~\ref{fig:ratiofit} in terms of the initial input parameters $\sqrt{\alpha}$ versus the corresponding $\sqrt{\alpha_{\mathrm{CT}}}$ or equivalently $\sqrt{J/K}$ versus $\sqrt{(J/K)_{\mathrm{CT}}}$,  quantities proportional to the domain-wall width in lattice units within the micromagnetic approximation. We find a unique correspondence between the microscopic and apparent continuum values. However, their deviation becomes particularly pronounced near the critical point $J/K=3/2$. Whereas continuum theory permits a wall width that decreases continuously to zero, the discrete lattice model undergoes a transition to an atomically sharp easy-axis wall at a finite exchange-to-anisotropy ratio. For broad domain walls, the difference $\sqrt{J/K}-\sqrt{(J/K)_\mathrm{CT}}$ tends to zero, demonstrating the recovery of the continuum limit for broad domain walls. The deviation
is approximately 10\% at about $\alpha = 4$, dropping to about 3\% at $\alpha = 10$.

Consider a typical experimental situation that the magnetic anisotropy $K$ is determined independently, for example, from magnetometry or ferromagnetic resonance, and the exchange interaction  $J$  had been determined  making use of the domain-wall profile, then the mapping between $\alpha_{\mathrm{CT}}$ and $\alpha$ in Fig.~\ref{fig:ratiofit} is particularly relevant. Assuming that the independently determined anisotropy agrees with the microscopic value, $K_{\mathrm{CT}}=K$, the continuum-derived exchange parameter satisfies $J_{\mathrm{CT}}/J=\alpha_{\mathrm{CT}}/\alpha$. Since $\alpha_{\mathrm{CT}}<\alpha$, fitting the domain-wall profile with the continuum solution systematically underestimates the exchange interaction.

\begin{figure}[t]
\centering
\includegraphics[width=0.86\columnwidth]{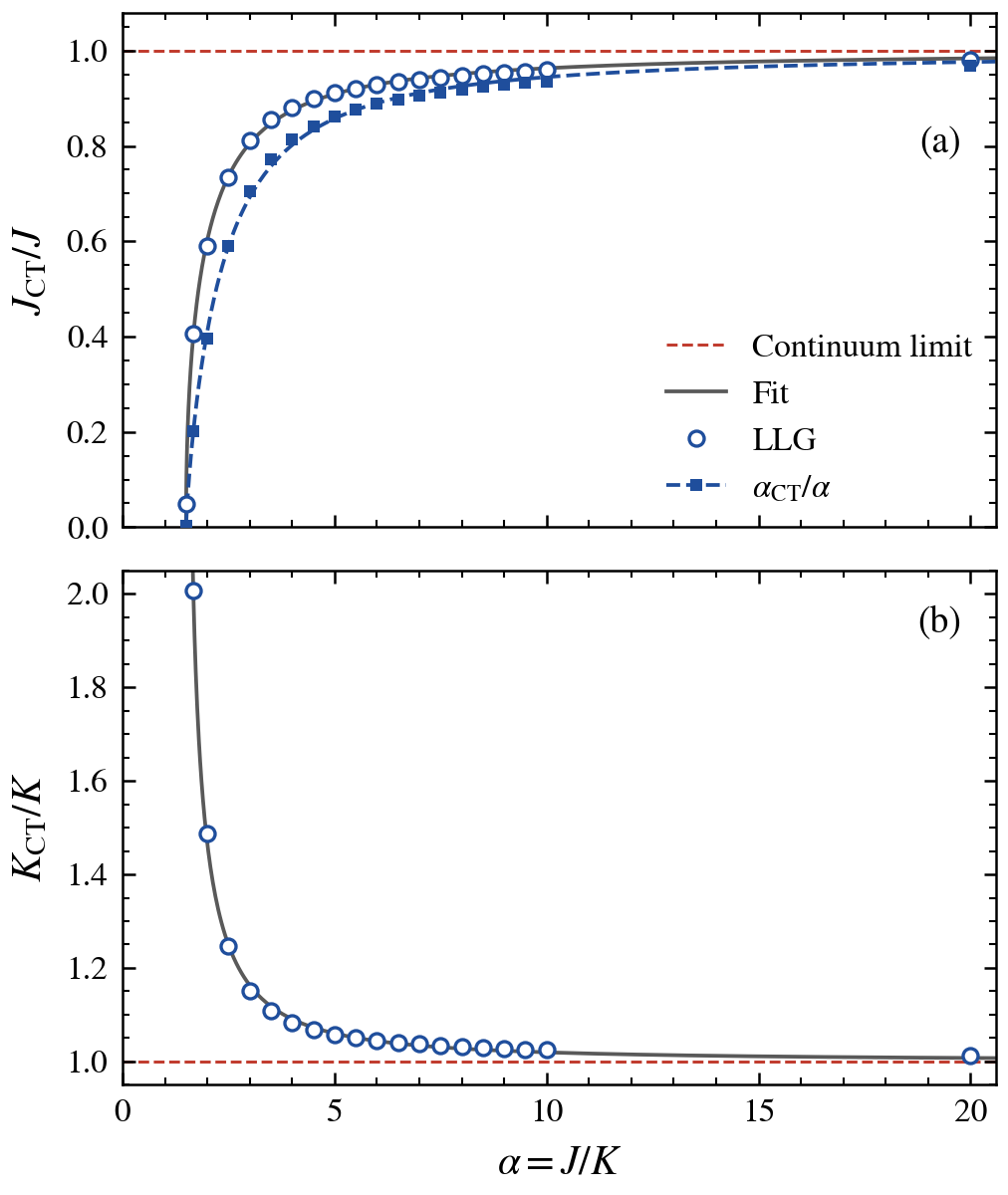}
\caption{Continuum parameters obtained from the fitted wall profile. Ratios (a) $J_{\rm CT}/J$ and (b) $K_{\rm CT}/K$ as functions of $\alpha=J/K$. The dashed lines mark the continuum limit. The additional dashed curve in panel (a) shows $\alpha_{\rm CT}/\alpha$, which equals $J_{\rm CT}/J$ when $K_{\rm CT}=K$. The fit underestimates $J$ and overestimates $K$, with a smaller relative deviation in $K$.}
\label{fig:parameterfit}
\end{figure}

A complementary theoretical comparison can be made by determining both continuum parameters from the domain-wall profile and its formation energy. The resulting fit $\alpha_\mathrm{CT}$ and the corresponding $\alpha$ obtained form Fig.~\ref{fig:ratiofit}  result into ratios $J_{\mathrm{CT}}/J$ and $K_{\mathrm{CT}}/K$ shown in Figs.~\ref{fig:parameterfit}(a) and~\ref{fig:parameterfit}(b), respectively. In this case, the continuum approximation underestimates the exchange interaction and overestimates the magnetic anisotropy, with the larger relative error occurring in $J$. Both ratios approach unity for broad domain walls, whereas the deviations increase rapidly near the atomically sharp-wall transition. Even at $\alpha=5$, well above the critical ratio $\alpha_\mathrm{c}=3/2$, the exchange interaction is underestimated by approximately $10$\% and the anisotropy overestimated by about $5$\%. Compared with the case of independently determined anisotropy, the overestimation of \(K\) in the two-parameter fit partially compensates for the underestimation of \(J\), resulting in a smaller relative error in the inferred exchange interaction. Thus, the systematic errors in the extracted interaction parameters also depend on the choice of experimental observables.

\begin{figure}[t]
\centering
\includegraphics[width=0.86\columnwidth]{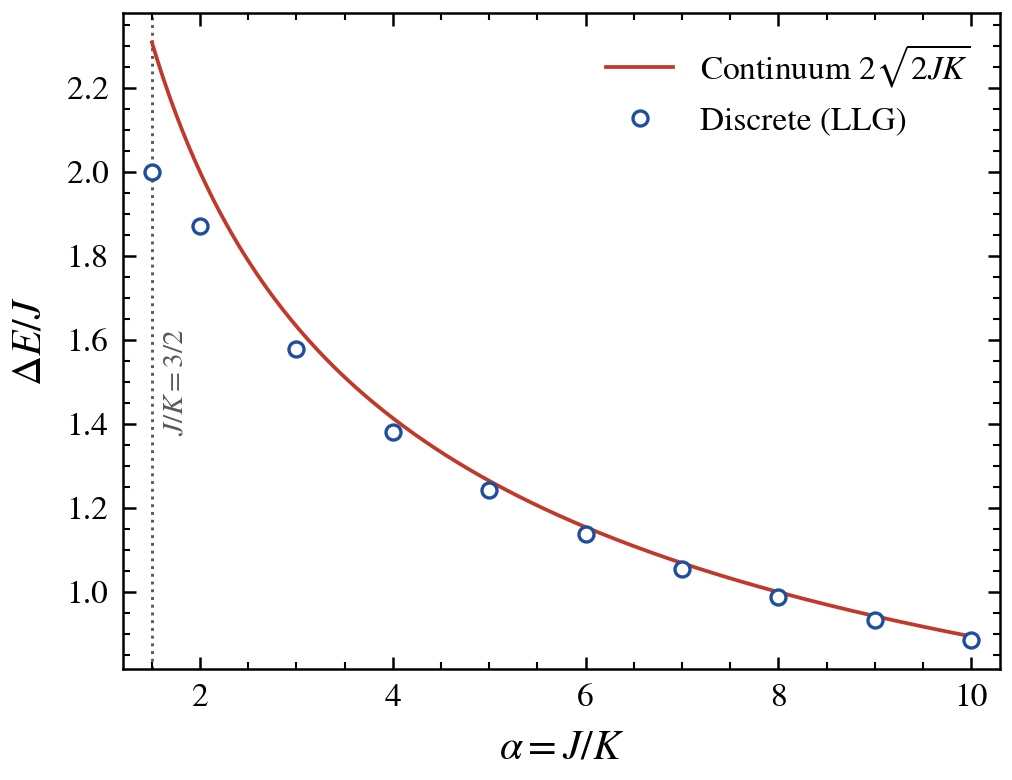}
\caption{Domain-wall energy relative to the ferromagnetic state at $D_y=0$. Circles show $\Delta E_\mathrm{lat}$ from the relaxed discrete chain and the solid line shows the continuum prediction $\Delta E_\mathrm{cont}=2\sqrt{2JK}$. The dotted line marks the discrete transition at $J/K=3/2$.}
\label{fig:energycomparison}
\end{figure}

The limitations of the continuum approximation can also be quantified by comparing domain-wall formation energies.  For the discrete lattice model, we calculate $\Delta E_{\mathrm{lat}}=E_{\mathrm{DW}}-E_{\mathrm{FM}}$ from the relaxed domain-wall configuration relative to the corresponding ferromagnetic chain with identical model parameters. Since a perfectly sharp, collinear domain wall has a formation energy of exactly $2J$, independent of the magnetic anisotropy $K$, the relaxed wall energy is bounded from above by $\Delta E_{\mathrm{lat}}\leq 2J$. In contrast, inserting Eq.~\eqref{eq:rel_micro-to-atomistic_parameters} into the continuum results \eqref{eq:micromagnetic_width} and \eqref{eq:micromagnetic_DW_energy} gives
\begin{equation}
\Delta E_\mathrm{cont}=2\sqrt{2JK}\quad\text{or}\quad 
\frac{\Delta E_\mathrm{cont}}{J}=\sqrt{\frac{8}{\alpha}}=4a\frac{1}{w}\, .
\label{eq:continuum_energy_lattice}
\end{equation}
As shown in Fig.~\ref{fig:energycomparison}, the continuum result lies above the discrete energy throughout the investigated range of $\alpha$ and diverges for decreasing domain walls $w$. The difference is largest near the atomically sharp-wall transition and decreases as the wall broadens. This is the energetic counterpart of the profile mismatch: whereas the continuum exchange energy assigns an increasingly large cost to rapid spatial variations of the magnetization, the exchange-energy cost of an abrupt reversal between neighboring lattice spins remains finite.

\section{Domain walls in the $J$--$K$--$D$ model}
Having established the transition to an atomically sharp easy-axis state (EAS) at $J/K=3/2$ in the absence of DMI, we now investigate how this transition is modifies by finite DMI. Within the continuum description, DMI selects the chirality of a N\'eel wall without changing its equilibrium profile or width. In the discrete spin-lattice model, however, DMI induces canting near the wall center, rendering the perfectly collinear EAS nonstationary.

We restrict our analysis to DMI strengths for which the ferromagnetic domains remain energetically favored over a homogeneous Dzyaloshinskii spin spiral. For a long-wavelength spiral the energy density relative to the ferromagnet is approximately
\begin{equation}
\delta e(q)\simeq \frac{J}{2}q^2-D_yq+\frac{K}{2},
\end{equation}
which yields the condition $JK>D_y^2$, or equivalently $(D_y/J)^2(J/K)<1$. All finite-DMI results presented below satisfy this approximate criterion. To distinguish stationarity from local stability, we examine both the residual torques and the Hessian of the magnetic energy. Vanishing torques, within the numerical tolerance of the LLG relaxation, establish stationarity, whereas a positive-definite Hessian establishes local stability. We first demonstrate that finite DMI removes the stationary EAS and then investigate the resulting changes in the Hessian spectrum, the stability and evolution of localized soft modes  and the DMI induced canting of the domain-wall profile. For convenience we place frequently the wall between sites $0$ and $1$ instead of $N/2$ and $N/2+1$.

\subsection{Absence of a  stationary easy-axis wall}
To understand why finite DMI eliminates the perfectly collinear EAS, we examine the generalized torque $\tau_i=-\partial E/\partial\varphi_i$ on each interior spin. Differentiating Eq.~\eqref{eq:spin-model_Ephi}, the stationary condition is
\begin{eqnarray}\label{eq:DMI-stationarity}
0=\tau_i
&=&J(\sin\Delta_{i-1}-\sin\Delta_i)  \\
&&-D_y(\cos\Delta_{i-1}-\cos\Delta_i)-K\sin(2\varphi_i)\, , \nonumber
\end{eqnarray}
where $\Delta_i=\varphi_{i+1}-\varphi_i$.

For the ideal EAS, all spins are aligned parallel or antiparallel to the easy axis. Consequently, the exchange and anisotropy torques vanish. The DMI torque, however, remains finite at the wall because $\cos\Delta_i$ changes from $+1$ within the ferromagnetic domains to $-1$ across the antiparallel bond. Introducing $\sigma_i:=\cos\Delta_i$, the torque evaluated at the ideal EAS is
\begin{equation}
\tau_i^{\mathrm{EAS}}
=-D_y(\sigma_{i-1}-\sigma_i),
\end{equation}
with
\begin{equation}
\tau_0^{\mathrm{EAS}}=-2D_y,\qquad
\tau_1^{\mathrm{EAS}}=+2D_y,
\label{eq:DMI-force}
\end{equation}
where sites $0$ and $1$ denote the two spins adjacent to the atomically sharp wall. The torque vanishes at all other sites.

Although the uniaxial anisotropy produces no torque at the perfectly collinear EAS, any canting away from the easy axis incurs a positive anisotropy energy proportional to $Ku_i^2$. The equilibrium canting therefore results from competition between the linear energy gain induced by DMI and the quadratic restoring energy associated with anisotropy and exchange, as quantified by the linear-response analysis below.

Thus, finite DMI generates equal and opposite torques on the two spins bordering the wall, rendering the perfectly collinear EAS nonstationary for any nonzero $D_y$, irrespective of $J/K$. This eliminates the collinear solution underlying the zero-DMI transition and suggests a crossover toward an increasingly localized, but still canted, domain wall. We examine this behavior through the Hessian spectrum and the evolution of the domain-wall profile in the following subsections.

\subsection{Hessian spectrum}
\begin{figure}[t]
\centering
\includegraphics[width=0.93\columnwidth]{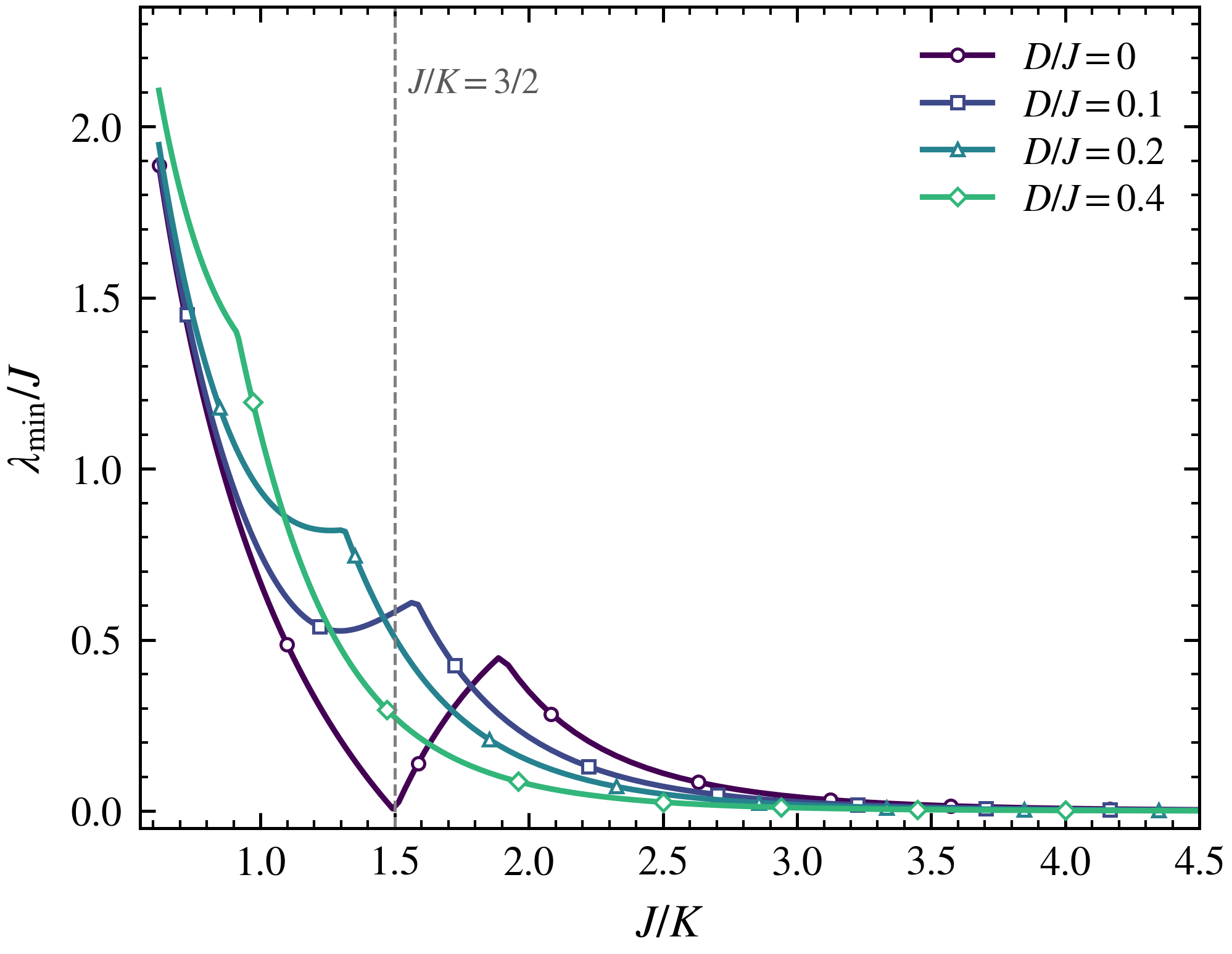}
\caption{Lowest eigenvalue of the in-plane Hessian at the relaxed configurations of a 100-spin chain, plotted against $J/K$ for several DMI strengths. At $D_y=0$, the localized wall mode vanishes at $J/K=3/2$; finite DMI lifts the zero. The kinks are exact crossings between the lowest symmetric and antisymmetric modes, not discontinuities of the relaxed wall.}
\label{fig:hessian}
\end{figure}

To investigate how finite DMI affects the local stability of the domain wall near the zero-DMI critical ratio $J/K=3/2$, we calculate the lowest eigenvalue of the Hessian $H_{ij}=\partial^2E/(\partial\varphi_i\partial\varphi_j)$ by numerical diagonalization at the relaxed spin configurations for different DMI strengths $D$ in units of the exchange interaction $J$. As shown in Fig.~\ref{fig:hessian}, the localized wall mode softens to zero at $J/K=3/2$ for $D_y=0$, signaling the loss of stability of the atomically sharp EAS toward finite spin canting in agreement with the analytical result derived in Sec.~\ref{sec:J-K-Model}.

At finite DMI, however, the situation changes qualitatively. The DMI-induced localized torque drives the spins away from their collinear easy-axis orientations and produces a finite homochiral canting of the spins near the wall center, while exchange and anisotropy provide the restoring stiffness. As a consequence, the lowest eigenvalue of the Hessian, evaluated at the relaxed canted configuration, remains positive for all finite DMI strengths for all finite ratios $J/K$ investigated, establishing the local stability of the relaxed canted-wall configurations. Calculations for chains containing 100 and 200 spins yield essentially identical results, demonstrating that the observed behavior is not a finite size artifact.

The kinks in Fig.~\ref{fig:hessian} arise from crossings between Hessian modes of different symmetry rather than discontinuities in the equilibrium spin configuration. For a domain wall centered between two lattice sites, the Hessian possesses mirror symmetry, allowing its eigenvectors to be classified by their parity about the central bond. Denoting the lowest symmetric and antisymmetric eigenvalues by $\lambda_{\mathrm S}$ and $\lambda_{\mathrm A}$, respectively, we have
\begin{equation}
\lambda_{\min}=\min(\lambda_S,\lambda_A)\, ,
\end{equation}
Since modes of opposite parity do not hybridize, their eigenvalues can cross, producing a kink in $\lambda_{\min}$ without a discontinuity in the domain-wall profile.

Appendix~\ref{app:EigV_Hessian} presents the symmetry-resolved spectra, the corresponding real-space eigenvectors, and the detailed localization analysis. Together with the absence of a stationary collinear EAS, the positive Hessian eigenvalues support the interpretation of the finite-DMI behavior as a crossover rather than a continuation of the zero-DMI transition.

\subsection{Minimum magnetization and crossover}

To characterize the evolution of the domain-wall profile across the crossover, we introduce the minimum absolute out-of-plane spin component,
\begin{equation}
m_{\min}=\min_i |S_{i,z}|\qquad i\in\{1,\cdots ,N\}\, ,
\label{eq:mmin}
\end{equation}
Unlike the conventional domain-wall width, which becomes ambiguous when the magnetization rotation is confined to only a few lattice sites, $m_{\min}$ remains well-defined and measures the departure from the perfectly collinear EAS. For an atomically sharp easy-axis wall, $m_{\min}=1$, whereas any finite canting gives $m_{\min}<1$. In our simulations, the wall is centered between two sites of a chain with $N=100$ spins, satisfying the symmetry relation $-S_{i,z}=S_{N+1-i,z}$. Consequently, the discrete profile need not contain a site with $S_{i,z}=0$, even when the corresponding continuum profile passes through zero at the wall center.

\begin{figure}[t]
\centering
\includegraphics[width=0.93\columnwidth]{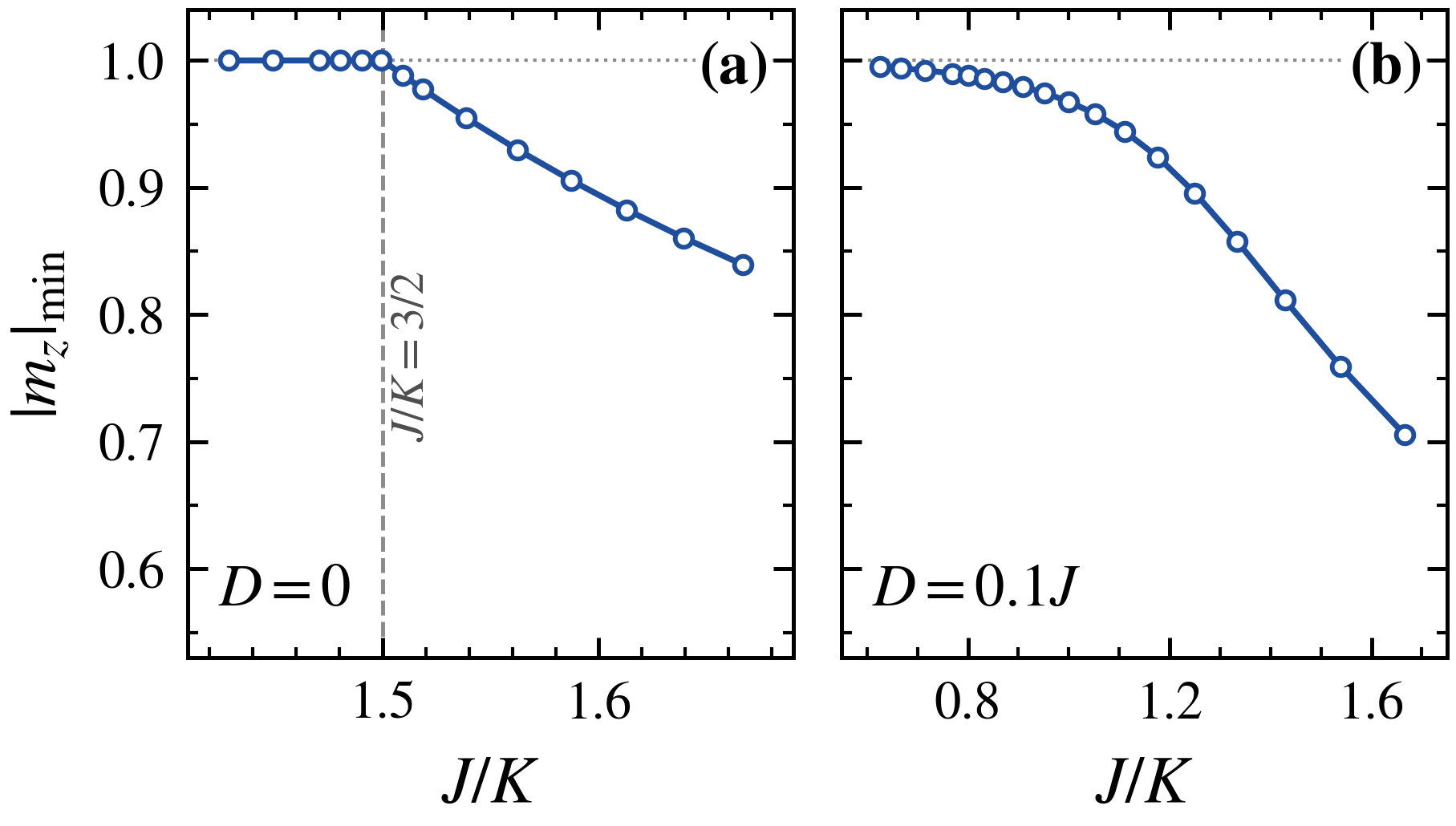}
\caption{Minimum absolute out-of-plane spin component $m_{\min}$ of relaxed domain-wall configurations as a function of $J/K$: (a) $D_y=0$ and (b) $D_y=0.1J$. In the absence of DMI, $m_{\min}$ reaches unity at $J/K=3/2$, signaling the atomically sharp EAS. At finite DMI it approaches unity continuously but remains below one for every finite anisotropy because the wall-adjacent spins remain canted.}
\label{fig:mmin}
\end{figure}

Figure~\ref{fig:mmin} illustrates the qualitatively different behavior with and without DMI. For $D_y=0$, $m_{\min}$ reaches unity at the critical ratio $J/K=3/2$ and remains exactly one in the EAS regime. At finite DMI, by contrast, $m_{\min}$ increases smoothly toward unity with increasing anisotropy but does not reach an exact plateau at any finite $J/K$. Thus, even when the magnetization reversal is strongly localized around the central bond, the spins adjacent to the wall retain a finite canting away from the easy axis.

The continuous evolution of $m_{\min}$, together with the absence of a stationary EAS and the lifting of the localized Hessian soft mode, supports the replacement of the zero-DMI sharp-wall transition by a crossover. The residual canting in the strongly anisotropic regime can be quantified analytically, as discussed in the following subsection.

\subsection{\label{ssec:LTEAS}Linear theory in the sharp-wall regime}

To calculate the small tilting angles in the sharp wall with
\begin{equation}
\varphi_i^0=
\begin{cases}
-\pi/2,&i\leq0,\\
+\pi/2,&i\geq1,
\end{cases}
\end{equation}
and each angle $\varphi$  is written as the strict up/down value  $\varphi_0$ plus a small angular deviation $u_i$:
\begin{equation}
\varphi_i=\varphi_i^0+u_i,\qquad u_i\ll1\, ,\qquad\varphi_i^0=\pm\frac{\pi}{2} .
\label{eq:linear-angle}
\end{equation}
Linearizing the stationary equation \eqref{eq:DMI-stationarity} around the sharp wall, EAS, leads to the linear-response equation, which relates the angular deviation $\mathbf{u}$ to the torque $\boldsymbol{\tau}^\mathrm{EAS}$:
\begin{equation}
\matr{H}\mathbf{u}=\boldsymbol{\tau}^\mathrm{EAS}.
\label{eq:linear-response}
\end{equation}
Here, $\boldsymbol{\tau}^\mathrm{EAS}$ follows Eq.~\eqref{eq:DMI-force}. Inside the domain, the angular deviation satisfies the recurrence relation
\begin{equation}
-Ju_{i-1}+(2J+2K)u_i-Ju_{i+1}=0\quad i\ne0,1\, ,
\label{eq:linear-recurrence}
\end{equation}
which has the recurrence form of the Chebyshev polynomials of the second kind in Eq.~\eqref{eq:app_recurrence}, with $\lambda=1+1/\alpha$ and $\alpha=J/K$.

At the two spins adjacent to the central bond,
the DMI produces finite driving terms:
\begin{align}
-Ju_{-1}+2Ku_0+Ju_1&=\phantom{-}2D_y,\\
Ju_0+2Ku_1-Ju_2&=-2D_y.
\label{eq:linear-boundary}
\end{align}
The DMI thus acts as a localized source term,
whereas exchange and anisotropy determine the
restoring stiffness of the spin configuration.
At the collinear EAS, the DMI contribution to the Hessian vanishes, whereas its first derivative produces the finite torques on the two spins adjacent to the wall. At the wall, the homogeneous recurrence is replaced by an inhomogeneous one with the inhomogeneity introduced by the DMI torque. Solving the linear-response equations applying the characteristic-root method gives  for \eqref{eq:linear-recurrence} the exponentially decaying solution characterized by
\begin{equation}
r=\frac{1+\alpha-\sqrt{1+2\alpha}}{\alpha},
\qquad \alpha=J/K,
\end{equation}
where $0<r<1$. Using $u_{-1}=ru_0$, $u_2=ru_1$, and the antisymmetry $u_1=-u_0$, the boundary equations yield for the angular deviation at the wall site,
\begin{equation}
u_0=-u_1=-\frac{2D_y}{2K-J(1+r)}
=-\frac{2D_y}{J}\frac{\alpha}{1-2\alpha+\sqrt{1+2\alpha}}\, .
\label{eq:linear-amplitude}
\end{equation}
This motivates a useful definition of the atomically narrow state at finite DMI. Here we use
\begin{equation}
\varphi_n=
\begin{cases}
-\pi/2-\theta r^{|n|},&n\leq0,\\
+\pi/2+\theta r^{n-1},&n\geq1,
\end{cases}
\label{eq:canted-sharp-wall}
\end{equation}
as the ansatz function of the canted sharp wall in the presence of DMI. The two spins adjacent to the central bond tilt by the same magnitude $\theta=-u_0$ toward a common in-plane direction selected by the sign of $D_y$, while the deviation decays into the domains with ratio $r$. Thus, for an atomically narrow wall the canting angle $\theta$, rather than a continuum width, characterizes the remaining internal structure.

The solution applies on the stable EAS side of the zero-DMI transition, $\alpha<\alpha_\mathrm{c}=3/2$, provided that the induced canting remains small. For $D_y>0$, the angular deviations satisfy $u_0<0$ and $u_1>0$, corresponding to canting of both wall-adjacent spins toward the negative $x$ direction and hence to clockwise rotation.
At the critical ratio $\alpha_\mathrm{c}=3/2$, the decay factor approaches $r=1/3$, corresponding to $\ell=a/\ln 3$. Thus, although the canting amplitude becomes strongly enhanced near the transition, its spatial extent remains of atomic scale. This demonstrates that the enhanced response originates from the softening of the localized Hessian mode rather than from a diverging spatial decay length.
For a finite chain, the same recurrence can be solved exactly using Chebyshev polynomials of the second kind, $U_n(x)$, with $x=1+1/\alpha$. The finite-chain solution approaches the exponentially decaying form derived above as the distance between the wall and the fixed boundaries increases.

The denominator of \eqref{eq:linear-amplitude} vanishes at $\alpha_\mathrm{c}=3/2$, reflecting the softening of the localized Hessian mode. Expanding around this critical ratio $\alpha_\mathrm{c}$ from below, $\alpha < \alpha_\mathrm{c}$ yields
\begin{eqnarray}
u_0\simeq &&-\frac{4D_y}{3K} \frac{1}{\alpha_\mathrm{c}-\alpha}\, , \qquad \alpha\lesssim\alpha_\mathrm{c},  \quad |u_0|\ll 1 \\
& = &-\frac{8D_y}{9(K-2J/3)},\qquad m_{\min}\simeq\cos u_0\simeq1-\frac{u_0^2}{2}\nonumber \, 
\label{eq:linear-critical}
\end{eqnarray}
showing that the canting amplitude is strongly enhanced as the critical ratio is approached from the EAS side.
The apparent divergence of the linear-response amplitude as $\alpha\to\alpha_\mathrm{c}$ signals the breakdown of the small-angle approximation rather than a divergence of the equilibrium canting angle. Accordingly, the small-angle approximation requires $|u_0|\ll 1$, which becomes increasingly restrictive near $\alpha_\mathrm{c}$. The strong response of the canting angle near $\alpha_\mathrm{c}$ shows that the wall sharpness changes continuously with $K$ and $D_y$. This differs fundamentally from the micromagnetic limit, where DMI does not change the wall profile. The sign of $D_y$ uniquely determines the sign of $u_0$ and hence the homochirality of the wall. For $D_y>0$, the negative value $u_0<0$ is consistent with clockwise rotation. In the strong-anisotropy regime, $\alpha < \alpha_\mathrm{c}$, linear response gives $1-|m_{\min}|\simeq\frac{u_0^2}{2}\approx\frac{1}{2}\left(\frac{8}{9}\right)^2\left(\frac{D_y}{J}\right)^2\left(1+\frac{\alpha}{\alpha_\mathrm{c}} + \left(\frac{\alpha}{\alpha_\mathrm{c}}\right)^2\right)$, which explains the behavior of Fig.~\ref{fig:mmin}.

\begin{figure}[t]
\centering
\includegraphics[width=0.93\columnwidth]{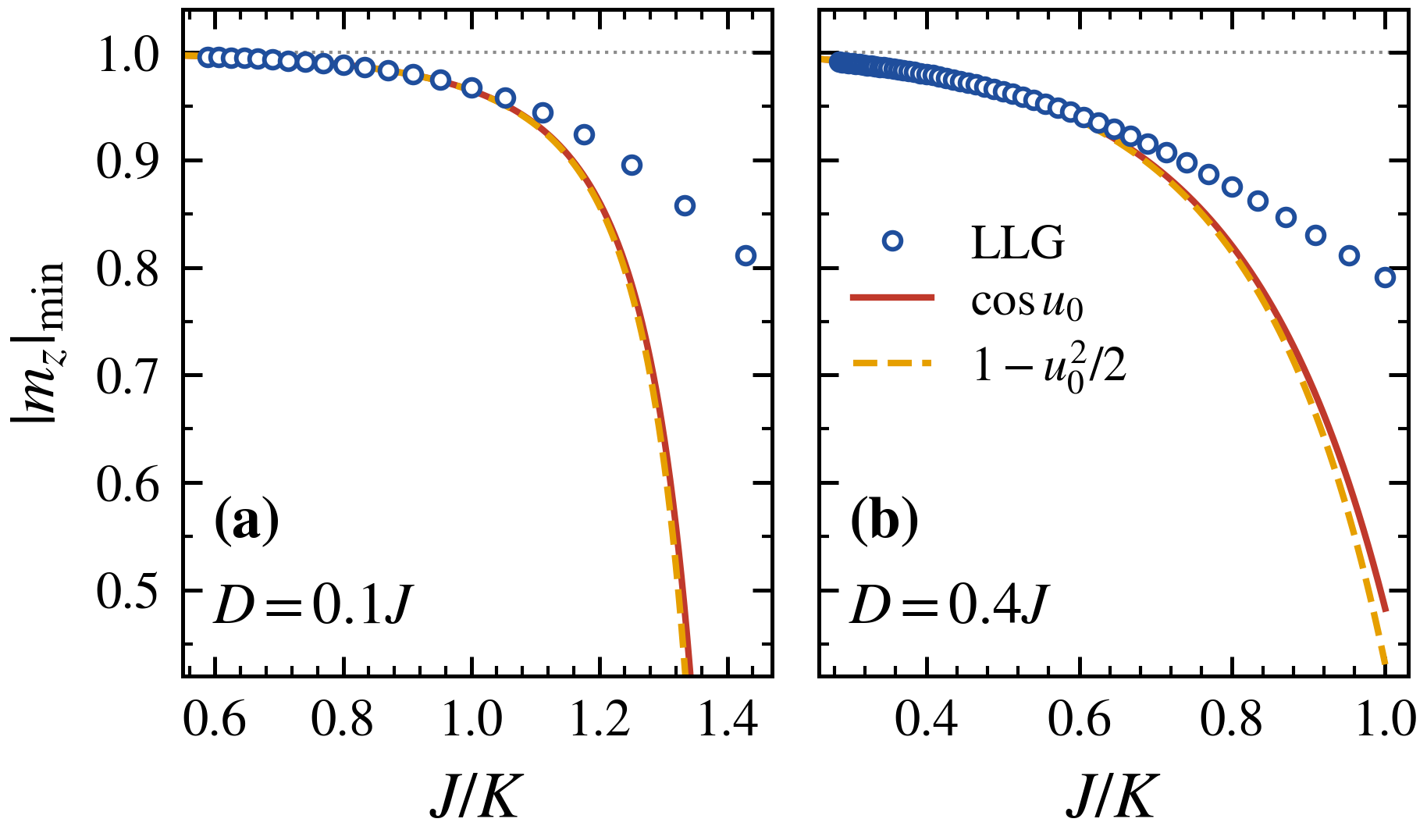}
\caption{Minimum magnetization from LLG relaxation (circles) and linear response for (a) $D_y=0.1J$ and (b) $D_y=0.4J$. The solid line uses $m_{\min}=\cos u_0$ and the dashed line its small-angle expansion $1-u_0^2/2$. Agreement between LLG and analytlical approximation improves as $\alpha=J/K$ decreases and the wall enters the strong-anisotropy regime.}
\label{fig:linear}
\end{figure}

Figure~\ref{fig:linear} compares the minimum absolute out-of-plane magnetization, $m_{\min}\simeq\cos u_0$, obtained from linear response with the results of the atomistic spin-dynamics simulations for two DMI strengths. Close to $\alpha_\mathrm{c}$, the analytical approximation becomes inaccurate because the canting is no longer sufficiently small. Agreement improves with increasing anisotropy (decreasing $\alpha$), as the wall approaches the strongly localized regime. For larger DMI, a stronger anisotropy is required to obtain comparable agreement. These results establish the range of validity of the linear theory and demonstrate that DMI produces a finite canting even in the strongly anisotropic limit.

\section{Discussion}
The discrete spin-lattice model reveals a fundamental distinction between domain walls with and without DMI. In the absence of DMI, the competition between ferromagnetic exchange and easy-axis anisotropy produces a continuous transition to an atomically sharp easy-axis state at $\alpha_\mathrm{c}=J/K=3/2$, as witnessed by the kink of $|m_\mathrm{min}|$ in Fig.~\ref{fig:mmin}(a). The transition is associated with the softening of a localized Hessian mode. Finite DMI, however, exerts a nonzero torque on the spins adjacent to the wall, removing the perfectly collinear state as a stationary solution. The resulting canting lifts the zero-mode instability along the relaxed configurations investigated and replaces the sharp-wall transition by a smooth crossover. By this we mean that, with increasing anisotropy, the domain-wall profile evolves gradually from an extended spin rotation toward an atomically localized but still canted configuration, without reaching a perfectly collinear state at finite anisotropy.

The physical origin of this behavior can be understood by analogy with the rounding of a continuous phase transition by an external field. Just as a magnetic field couples linearly to the order parameter of a ferromagnet, DMI couples linearly to the localized canting mode of the domain wall, thereby removing the zero-canting stationary configuration. Thus, the signed canting amplitude $u$ of the spins adjacent to the wall serves as a local order parameter: it vanishes in the perfectly collinear easy-axis state and becomes finite in the canted-wall configuration. Whereas the minimum absolute out-of-plane magnetization satisfies $1-|m_{\min}|\simeq u^2/2$ and therefore measures only the magnitude of the canting, $u$ also distinguishes its two symmetry-related orientations. Near $\alpha_\mathrm{c}$, the restoring stiffness of this mode vanishes in the absence of DMI, giving rise to a strongly enhanced response to even a weak DMI-induced torque. The divergence predicted by linear response signals the breakdown of the small-angle approximation rather than a physical divergence. Remarkably, the spatial decay length of the linear-response deformation remains finite at $\alpha_\mathrm{c}$, demonstrating that the enhanced susceptibility is associated with a localized instability rather than a diverging domain-wall width. This behavior differs fundamentally from the continuum description, in which DMI selects the chirality of a Néel wall without changing its equilibrium profile or width.

These findings also have implications for extracting magnetic interaction parameters from measured domain-wall profiles. When the magnetic anisotropy is determined independently, fitting an atomically narrow domain-wall profile with the conventional micromagnetic solution systematically underestimates the exchange interaction. If both exchange and anisotropy are instead inferred from the domain-wall profile and its formation energy, the continuum approximation underestimates the former and overestimates the latter, with the overestimated anisotropy partially compensating for the error in the exchange interaction. Thus, the systematic errors in the extracted interaction parameters also depend on the choice of experimental observables. Importantly, appreciable deviations already occur for parameter spaces of $J$ and $K$ before the atomically sharp-wall regime is reached. Finite DMI does not restore the validity of the continuum approximation, even though it prevents the formation of an exactly collinear sharp wall. Consequently, atomically narrow domain walls should be analyzed using a discrete spin model, particularly in ultrathin magnetic films and low-dimensional magnetic materials where exchange, anisotropy, and DMI can be comparable.

The pronounced sensitivity of the domain-wall profile to anisotropy and DMI near $\alpha_\mathrm{c}=3/2$ originates from the softening of a localized Hessian mode. At this critical ratio, the exchange-driven tendency to spread the magnetization reversal over several lattice sites is exactly balanced, to quadratic order, by the restoring contribution of the easy-axis anisotropy. Consequently, the energy cost of a localized deformation vanishes, and even a weak DMI-induced torque produces a strongly enhanced canting of the spins adjacent to the wall. Within linear response, the canting amplitude scales as $|u_0|\propto |D_y|/(\alpha_\mathrm{c}-\alpha)$, demonstrating the increasing susceptibility of the wall as the critical point is approached from the EAS side. The apparent divergence signals the breakdown of linear response, where higher-order terms stabilize a finite deformation. Finite DMI thus acts as a symmetry-breaking field conjugate to the localized wall deformation, replacing the zero-DMI instability by a smooth crossover between extended and strongly localized domain walls.

The $J$--$K$ and $J$--$K$--$D$ models therefore have qualitatively different transition behavior. At $D_y=0$, the easy-axis state is stationary, and its loss of stability is accompanied by a zero Hessian eigenvalue at $J/K=3/2$. Finite DMI forces the two wall-adjacent spins to cant. That torque remains nonzero at arbitrarily large $K$, so the exact easy-axis state is replaced by a canted state and the zero Hessian mode is lifted. Experimentally, the predicted distinction can be tested by resolving whether the central spins become fully easy-axis aligned or retain a small, chirality-locked transverse component as the wall narrows.

This distinction matters for the definition of the wall width. A continuum theory is useful for a smooth profile, but it becomes ambiguous when the wall length is only two or three sites. The lattice observable $m_{\min}$ remains well defined in that regime and distinguishes an easy-axis state from a DMI-canted wall. The Hessian provides an independent stability indication, while the linear formula for the magnetization identifies the asymptotic behavior toward the sharp limit.

The discussion throughout this paper is based on the $J$--$K$ and $J$--$K$--$D$ models with nearest-neighbor exchange interactions and DMI. In Appendix~\ref{app:nnn_critical}, we extend the analysis of the $J$--$K$ model to include next-nearest-neighbor exchange, described by the $J_1$--$J_2$--$K$ model. We consider ferromagnetic chains with $J_1>0$ and $J_2\geq-J_1/4$, for which the ferromagnetic state remains an exchange ground state. The underlying physics is preserved: the atomically sharp easy-axis state (EAS) loses stability through the softening of a localized, reflection-antisymmetric Hessian mode, giving rise to a continuous transition toward a canted domain-wall state (DWS). However, the critical ratio cannot be obtained by simply replacing $J$ with $J_1+4J_2$, as might be suggested by the micromagnetic exchange stiffness $\Ab=a(J_1+4J_2)/2$. Instead, the critical anisotropy depends on the discrete exchange bonds crossing the domain wall. For small $|J_2|/J_1$, we obtain
\begin{equation}
K_\mathrm{c}\simeq\frac{2}{3}J_1+\frac{32}{81}J_2,
\label{eq:discussion_Kc}
\end{equation}
or, equivalently,
\begin{equation}
\alpha_\mathrm{c}\simeq\frac{3}{2}\left(1-\frac{16}{27}\frac{J_2}{J_1}\right),
\label{eq:discussion_alphac}
\end{equation}
as derived in Eqs.~(\ref{eq:nnn_K_expansion}) and (\ref{eq:nnn_alpha_expansion}). Thus, competing antiferromagnetic next-nearest-neighbor exchange ($J_2<0$) reduces the anisotropy required to stabilize the EAS up to about 10\%, whereas ferromagnetic $J_2>0$ increases it. Notably, the critical anisotropy remains finite even as the micromagnetic exchange stiffness vanishes at $J_2=-J_1/4$, highlighting the limitations of the continuum description for atomically sharp domain walls.

The parameter extraction also indicates a practical limit on micromagnetic fitting. When DMI is zero, the continuum formula agrees with the atomistic simulations only away from the easy-axis transition. Finite DMI does not restore the validity of the smooth-profile assumption; the crossover can still occur within a few lattice spacings. Measurements of atomically narrow walls should therefore be interpreted with a discrete model whenever the fitted width is comparable to the lattice constant.

\section{Summary and Conclusions}
We have investigated the structure and stability of narrow magnetic domain walls in one-dimensional $J$--$K$ and  $J$--$K$--$D$ spin-lattice models with ferromagnetic exchange, uniaxial anisotropy, and Dzyaloshinskii–Moriya interaction, and compared their behavior with the conventional micromagnetic description. In the absence of DMI, the discrete $J$--$K$ model exhibits a continuous transition to an atomically sharp easy-axis wall at the critical ratio $\alpha_\mathrm{c}=J/K=3/2$, characterized by the softening of a localized Hessian mode. Finite DMI, described in the $J$--$K$--$D$ model, removes the perfectly collinear wall as a stationary configuration and produces a smooth crossover toward a strongly localized but canted wall. The DMI-induced canting is calculated by means of atomistic spin-dynamics simulations and  can be described analytically in the strong-anisotropy regime, while its enhanced canting response near $\alpha_\mathrm{c}$ reflects the vanishing stiffness of the localized mode.

Our results establish the limitations of the continuum approximation for domain walls whose width approaches the atomic lattice spacing. Beyond the description of atomically sharp domain walls, our results reveal systematic errors in the magnetic interaction parameters extracted from domain-wall profiles using continuum theory, with the exchange interaction being underestimated even before the atomically sharp-wall regime is reached. The discrete description is therefore essential for a quantitative interpretation of atomically narrow domain walls, especially in ultrathin magnetic films, magnetic nanostructures, and two-dimensional magnetic materials, where the competition between exchange, anisotropy, and DMI can strongly influence the wall profile.

The pronounced sensitivity of atomically narrow domain walls to anisotropy and DMI near the critical exchange-to-anisotropy ratio suggests opportunities for controlling their magnetic structure through interface engineering, strain, or electric fields, particularly in ultrathin films and two-dimensional magnets.

\begin{acknowledgements}
S.B.\ acknowledges Markus Hoffmann for his contribution during the early stage of this work. G.L.\ gratefully acknowledges financial support by Outstanding Doctoral Students Overseas Study Program of the University of Science and Technology of China. S.B.\ gratefully acknowledges financial support by the European Research Council grant 856538 (project "3D MAGIC") and the Deutsche Forschungsgemeinschaft (DFG) through CRC~1238 (project number 277146847, subproject C01), SPP~2558 (project number 581438540, subproject BL444/18-1).
\end{acknowledgements}

\appendix

\section{Analytical determination of the critical
exchange-to-anisotropy ratio}
\label{app:critical}

In this Appendix, we derive the critical ratio
$\alpha_\mathrm{c}=J/K$ at which the atomically sharp
easy-axis state loses stability in the
thermodynamic limit. The derivation is based
on the Hessian of the discrete $J$--$K$
model and exploits the tridiagonal structure
of the resulting matrix.

We consider a chain of $N$ classical spins
with nearest-neighbor ferromagnetic exchange $J>0$, absence of DMI $D_y=0$, and uniaxial anisotropy $K>0$.
The energy \eqref{eq:spin-model_Ephi} simplifies to 
\begin{equation}
E[\varphi]=-J\sum_{i=1}^{N-1}\cos(\varphi_{i+1}-\varphi_i)
-K\sum_{i=1}^{N}\sin^2\varphi_i ,
\label{eq:app_energy}
\end{equation}
and is solved subject to the twist boundary conditions
$\varphi_1=-\pi/2$ and $\varphi_N=\pi/2$.

For even $N$, the atomically sharp
easy-axis state consists of two oppositely
magnetized domains separated by a single
bond between sites $N/2$ and $N/2+1$.
The spins are collinear within each domain,
and their orientations differ by $\pi$
across the central bond.

\subsection{$N=4$ sites}\label{A_ssec:4-states}

By examining the case of the four-spin system ($N=4$), minimizing Eq.\ \eqref{eq:spin-model_Ephi} with the energy $E=J(\sin\varphi_2-\cos(\varphi_2-\varphi_3)-\sin\varphi_3)-K(\sin^2\varphi_2+\sin^2\varphi_3)$  with respect to the angles $\varphi_{2,3}$ and considering the symmetry condition $\varphi_2=-\varphi_3$ we arrive a the condition 
\begin{equation}
\frac{1}{2} \frac{\alpha}{1-\alpha}\cos\varphi_{2,3}=\sin\varphi_{2,3}\cos\varphi_{2,3}\quad\text{with}\quad\alpha = J/K\, .
\end{equation}
It can easily be seen that there is a continuous (2nd order) phase transition at the critical value of $\alpha_\mathrm{c}(4)= 2$. For $\alpha_\mathrm{c}(4) >2$, the two mobile spins are oriented at angles $\varphi_{2,3}=\arcsin[\alpha/[2(1-\alpha)]$ between $\pm \pi/2$ with the horizontal $x$-axis, but below $\alpha_\mathrm{c}(4) =  2$, they point in the down or up direction. We denote these states as the domain wall state (DWS) and the atomically sharp DW or the easy axis state (EAS), respectively. We note, that the EAS is not part of the continuum solution \eqref{eq:micromagnetic_DW-profile} and is a result of the discrete model. The definition of $\alpha$ relates to the definition of $\xi^2$ above, and reminds on an area.

\subsection{Hessian of the easy-axis state}

The Hessian  $\matr{H}$ is defined by
\begin{equation}
 H_{ij}=
 \left.
 \frac{\partial^2 E}
 {\partial\varphi_i\partial\varphi_j}
 \right|_{\mathrm{EAS}} .
 \label{eq:app_H}
\end{equation}
The anisotropy contributes $2K$ to each
diagonal element. Each ferromagnetically
aligned bond contributes $J$ to the
diagonal elements and $-J$ to the
corresponding off-diagonal elements.
The antiparallel central bond contributes
with the opposite sign.

Because the easy-axis state is symmetric
under reflection about the central bond,
the instability can be analyzed in the
reflection-symmetric sector. Introducing
$n=N/2-1$ independent angular deviations,
the reduced Hessian takes the form
\begin{equation}
  \matr{H}_n(\alpha)=J
 \begin{pmatrix}
 2+2/\alpha & -1 & & &\\
 -1 & 2+2/\alpha & -1 & &\\
 &\ddots&\ddots&\ddots&\\
 &&-1&2+2/\alpha&-1\\
 &&&-1&-1+2/\alpha
 \end{pmatrix},
 \label{eq:app_DW-Hessian}
\end{equation}
where $\alpha=J/K$.

We introduce 
\begin{equation}
 \lambda=1+\frac{1}{\alpha}\, . 
 \label{eq:app_lambda_first}
\end{equation}
and the dimensionless eigenvalue $\epsilon=E/J$

The critical point $\lambda_\mathrm{c}$  is determined by the
vanishing of the lowest eigenvalue, or equivalently by
\begin{equation}
 \det\matr{H}_n(\lambda_\mathrm{c})=0.
 \label{eq:app_critical_condition}
\end{equation}
While we are in general only interested in the eigenvalue $E_m=\lambda_\mathrm{c}$ the characteristic eigenvalue equation is 
\begin{equation}
(\matr{H}_n/J- \epsilon_m \matr{1}_n)\mathbf{u}^{(m)}=0
\label{eq:app_H-Eigenvalue-problem}
\end{equation}
with solution found as in \eqref{eq:app_critical_condition} introducing the
dimensionless spectral parameter
\begin{equation}
 x=\lambda-\frac{\varepsilon}{2},
 \qquad E=J\varepsilon,
\end{equation}

\subsection{Uniform chain}\label{ssec:app_uniform_chain}
In the limit of the uniform chain, the wall defect represented by matrix element $H_{nn}$ in the Hessian \eqref{eq:app_DW-Hessian}, with $n=N/2-1$ disappears and the Hessian simplifies to 
\begin{equation}
 \matr{H}_n(2\lambda)= J\, \matr{T}_n(2\lambda)
 \label{eq:app_Chain-Hessian}
\end{equation}
with $n=N-2$ degrees of freedom and 
$\matr{T}$ is the  uniform tridiagonal matrix discussed in Appendix~\ref{app:Tridiagonal_Matrix}. This  Hessian \eqref{eq:app_Chain-Hessian} holds also true for the Hessian of the full $K$--$J$--$D$ model as the  DMI does not contribute to the Hessian at the homogeneous ferromagnetic configuration, however, it can contribute to the first derivative of the energy.

\subsection{Uniform chain spectrum}
We introduce a one-dimensional wave vector $q_m=\tau_m/a$,  with $q_m$ taken from the one-dimensional Brillouin zone (1DBZ). By means of \eqref{eq:app_eigenvalue_epsilon} the Hessian of the ferromagnetic chain 
yields the eigenvalue spectrum 
\begin{equation}
 E_m =2K+2J[1-\cos(q_m a)]\, ,
 \label{eq:app_uniform_spectrum}
\end{equation}
where
\begin{equation}
 q_m a=\frac{m\pi}{n+1},
 \qquad m=1,\ldots,n.
\end{equation}
All eigenvalues are positive for $J,K>0$, reflecting the local stability of the uniform easy-axis ferromagnetic state. In the thermodynamic limit,  $n\rightarrow \infty$, the lowest eigenvalue approaches  $E_{\min}(\mathrm{FM})=2K$ at $q=0$ and the highest approaches $E_{\min}(\mathrm{AFM})=2K+4J$ for the antiferromagnetic state at $q=\pi/a$. Thus, the eigenvalue spectrum of the uniform chain lies between 
\begin{equation}
  2K\le E_m \le 2K+4J
  \label{eq:app_chain_spectrum}
\end{equation}
The uniaxial anisotropy therefore opens a finite excitation gap, while the exchange interaction determines the dispersion above this gap. Since the spectrum remains positive for $J,K>0$, the uniform ferromagnetic state is locally stable.

In order to physically interprete the eigenvalue spectrum, we go one step back and look at the equation of motion of classical spins, \begin{equation}
\diff\vc{S}/\diff t=-\gamma \vc{S}\times \diff {E}/\diff \varphi\, ,
\label{eq:app_Blochdyn} 
\end{equation}
$\gamma$ is the gyromagnetic ratio, making the ansatz $\vc{S}(t)=\vc{S}\exp{(-i\omega t)}$, then in the  classical spin dynamics given by spin-lattice model~\eqref{eq:spin-model_Ephi} the eigenfrequency $\omega$ is related to the derivative of the energy with respect to the angle,
\begin{equation}
dE/d{\varphi_i}= -J\sin(\varphi_{i-1}-\varphi_i) + J\sin(\varphi_{i}-\varphi_{i+1}) -K\sin(2\varphi_i)
\end{equation}
for all $i=2,\dots, N-1$. Assuming small angle exitations around $\varphi=\pi/2$, \textit{i.e.}\ $\sin(\varphi_i-\varphi_{i+1}) \simeq \varphi_i-\varphi_{i+1}$,  $\sin(2\varphi)\simeq -2\varphi$ and $S_z\simeq |\vc{S}|$, we find that 
\begin{equation}
|\vc{S}|dE/d\grbf{\varphi} = \omega \mathbb{1} \,\grbf{\varphi}= J S \matr{T}_{N-2}(2+2/\alpha)\,\grbf{\varphi}\, ,
\end{equation}
with $\grbf{\varphi}$ being the $N-2$ dimensional state vector $\grbf{\varphi} := (\varphi_2, \dots , \varphi_{N-1})$. The eigenfrequencies  are then determined by the condition $\det{ \matr{T}_{N-2}(2+2/\alpha-\omega/(JS))}=0$, which is according to \eqref{eq:app_eigenvalue_epsilon} 
\begin{equation}
\omega(q) = 2KS+2JS(1-\cos q_m a)\, ,
\label{eq:spinwavedispersion}
\end{equation}
which is the well-known spin-wave spectrum of classical spins. 

\subsection{Hessian for chain with defect}
Having established the properties of the uniform ferromagnetic chain, we now turn to the domain wall, which can be regarded as a localized defect in this reference system. Since the defect modifies only a single diagonal element of the Hessian, its effect on the stability of the chain can be determined analytically from the known properties of the uniform tridiagonal matrix. Specifically, the Hessian in \eqref{eq:app_DW-Hessian} differs from $J\matr{T}_n(2\lambda)$ only in its last diagonal element and can be written as
\begin{equation}
\matr{H}_n(\lambda)/J
=\matr{T}_n(2\lambda)-3\mathbf{e}_n\mathbf{e}_n^\mathrm{T},
\end{equation}
where $\mathbf{e}_n=(0,\ldots,0,1)^\mathrm{T}$ is the $n$th unit vector.

Since the perturbation affects only the last diagonal element, the determinant of the domain-wall Hessian follows directly from a cofactor expansion,
\begin{align}
\frac{\det\matr{H}_n}{J^n}
&=\det\matr{T}_n(2\lambda)
-3\det\matr{T}_{n-1}(2\lambda)
\nonumber\\
&=U_n(\lambda)-3U_{n-1}(\lambda).
\label{eq:app_defect_det}
\end{align}
Here, the second determinant is the principal minor obtained by removing the last row and column of the uniform tridiagonal matrix. The condition for a vanishing eigenvalue, $\det\matr{H}=0$, of the domain-wall Hessian therefore becomes
\begin{equation}
U_n(\lambda)=3U_{n-1}(\lambda).
\label{eq:app_Cheb_condition}
\end{equation}

\subsection{Thermodynamic limit}

The atomically sharp domain-wall introduces a localized modification of the Hessian, corresponding to the reversed exchange bond at the wall center. This defect corresponding to a large excitation energy, which lies outside the spectrum of the energy dispersion \eqref{eq:app_uniform_spectrum} of the uniform chain. 

Consequently,  for positive $J$ and $K$, $\lambda=1+1/\alpha > \cos\tau$, which is outside the representation domain of $U_n(\lambda)$, with $|\lambda|\le 1$.  Unlike the uniform matrix, this perturbed matrix can support an eigenvector localized near the defect representing the EAS, if we extend the solution \eqref{eq:Cheby2nd_representation} to $|\lambda|>1$ by the transformation  $\lambda \rightarrow i\lambda$, and  
\begin{equation}
\lambda(\alpha)=1+\frac{1}{\alpha}=\cosh\tau(\alpha)\, . 
\label{eq:app_tau}
\end{equation}
Using the hyperbolic representation
of the Chebyshev polynomials,
\begin{equation}
 U_n(\cosh\tau)
 =\frac{\sinh[(n+1)\tau]}{\sinh\tau},
 \label{eq:app_hyperbolic}
\end{equation}
the condition for zero determinant  \eqref{eq:app_Cheb_condition} becomes
\begin{equation}
 \sinh[(n+1)\tau]
 =3\sinh(n\tau)\, .
 \label{eq:app_hyper_condition}
\end{equation}
Dividing by $\sinh(n\tau)$ and using the addition theorem yields
\begin{equation}
 \cosh\tau+\coth(n\tau)\sinh\tau=3.
 \label{eq:app_finite_n}
\end{equation}
In the thermodynamic limit, $n\rightarrow\infty$, we have $\coth(n\tau)\rightarrow1$, and therefore
\begin{equation}
 e^\tau=3,
 \qquad
 \cosh\tau=\frac53\, .
 \label{eq:app_tau_c}
\end{equation}

Combining this result with \eqref{eq:app_tau}, replace $\lambda$ by $x=\lambda-\epsilon/2$, and express the energy by the dimensionless value $\epsilon$ as $E=J\epsilon$ we obtain in analogy to \eqref{eq:app_uniform_spectrum}
\begin{eqnarray}
 E_\mathrm{EAS} &=&2K+2J[1-\cosh(\tau)]=2K-\frac{4}{3}J\nonumber \\
 &=& 2J\left(\frac{1}{\alpha}-\frac{2}{3}\right)\, .
 \label{eq:app_DW-energy}
\end{eqnarray}
The localized eigenvalue lies below the lower edge of the uniform-chain spectrum, $E_{\min}(\mathrm{FM})=2K$, by $4J/3$. However, its sign determines the local stability of the atomically sharp easy-axis wall.
For $\alpha=J/K<3/2$, the atomically sharp easy-axis state is stable against small angular deviations. For $\alpha=J/K<3/2$, the eigenvalue remains positive, and the wall is stable against small angular deviations. At the critical ratio
\begin{equation}
 \alpha_\mathrm{c}(\infty)
 =\left(\frac{J}{K}\right)_\mathrm{c}
 =\frac32\, , 
 \label{eq:app_alpha_critical}
\end{equation}
the eigenvalue of this localized mode vanishes. signaling the softening of a domain-wall mode. For $\alpha>3/2$, the eigenvalue becomes negative, and the atomically sharp wall is unstable toward a configuration with finite spin canting. This establishes the existence of a finite critical exchange-to-anisotropy ratio in the thermodynamic limit.

\subsection{Localization of atomically sharp wall}
At the critical point $\lambda_\mathrm{c}$ or $\alpha_\mathrm{c}$, respectively, the corresponding eigenvalue vanishes:
\begin{equation}
\matr{H}_n \mathbf{u}=0\, .
\end{equation}
Away from the defect, the eigenvector therefore satisfies exactly the same recursion relation \eqref{eq:app_eigenvector_recursion} of the uniform chain.  For $\lambda>1$, we introduced in \eqref{eq:app_tau} $ \lambda_\mathrm{c}=\cosh\tau_\mathrm{loc}$ and the eigenvector is given in analogy to \eqref{eq:app_eigenvector_chain} 
\begin{eqnarray}
u_{i+1}^{(\mathrm{loc})}&=&u_{1}U_i(x_\mathrm{loc})=u_{1}U_i(\cosh\tau_m) \nonumber \\
&=& u_{1}\frac{\sinh{(i+1)\tau_\mathrm{loc}}}{\sinh{\tau_\mathrm{loc}}}
\label{eq:app_eigenvector_EAS}
\end{eqnarray}
with the general solution 
\begin{equation}
u_i^{(\mathrm{loc})}=Ae^{i\tau_\mathrm{loc}} + Be^{-i\tau_\mathrm{loc}}
\end{equation}
In a semi-infinite chain, choosing the solution that decays exponentially away from the wall gives 
\begin{equation}
u_d\propto e^{-d\tau_\mathrm{loc}}\, ,
\end{equation}
where $d$ is the distance  counted in  the number of lattice spacings from the wall. According to \eqref{eq:app_tau_c}, $\tau_\mathrm{loc}=\ln 3$ and hence $u_d\propto 3^{-d}$. Thus, the amplitude of the critical eigenmode decreases by a factor of three for each lattice spacing away from the wall. The corresponding localization length  of the critical mode thst follows from $\tau=\ln 3$ is
\begin{equation}
\ell_\mathrm{loc}=a/\ln 3 \simeq 0.91~a \, .
\end{equation}

Thus, the instability at $J/K=3/2$ is driven by a mode strongly localized around the antiparallel bond, rather than by the softening of an extended spin-wave excitation if the chain mode. In other words, the uniform ferromagnetic chain remains stable throughout the transition, because its spectral gap \(2K\) never closes. It is the domain-wall defect that creates a localized mode within this gap, whose eigenvalue eventually crosses zero. This explains why the critical ratio remains finite in the thermodynamic limit and why the transition is an intrinsic consequence of the discrete spin-lattice model. 

For a finite chain with a fixed boundary, the exact zero-mode profile is instead proportional to $\sinh(j\tau)$, with $j$ measured from that boundary. It approaches the exponential form when expressed in terms of distance from the wall and the chain becomes long.

\section{Critical anisotropy for nearest- and next-nearest-neighbor exchange}
\label{app:nnn_critical}

We extend the stability analysis of Appendix~\ref{app:critical} to a one-dimensional ferromagnet with nearest-neighbor and next-nearest-neighbor exchange,
\begin{equation}
 E[\mathbf S]=
 -J_1\sum_i\mathbf S_i\cdot\mathbf S_{i+1}
 -J_2\sum_i\mathbf S_i\cdot\mathbf S_{i+2}
 -K\sum_i S_{i,z}^2,
 \label{eq:nnn_energy}
\end{equation}
where $J_1, K>0$, $J_2\geq-J_1/4$ and $|\mathbf S_i|=1$. Thus, we treat the entire ferromagnetic exchange regime including antiferromagnetic next-nearest-neighbor coupling, by an exact localized-mode secular equation (see also Appendix~\ref{app:EigV_Hessian}). We determine the anisotropy at which the collinear easy-axis state (EAS), with its magnetization reversing between two adjacent sites, loses local stability in the thermodynamic limit. The tridiagonal Hessian of Appendix~\ref{app:critical} is replaced by a pentadiagonal matrix with a localized defect involving three exchange bonds. 

\subsection{Pentadiagonal Hessian and the wall defect}
\label{app:nnn_hessian}

We follow Sect.\ref{ssec:LTEAS} and calculate the small tilting angles in the sharp wall, putting the wall between site 0 and 1 in the center of chain of $N$ sites. For planar spins $\mathbf S_i=(\cos\varphi_i,0,\sin\varphi_i)$, the EAS is specified by
\begin{equation}
 \varphi_i^{(0)}=
 \begin{cases}
 -\pi/2,&i\leq 0,\\
 +\pi/2,&i\geq 1.
 \end{cases}
 \label{eq:nnn_eas}
\end{equation}
Writing $\varphi_i=\varphi_i^{(0)}+u_i$, the quadratic energy variation
is $\delta^2 E=\frac12\mathbf u^{\mathsf T}H\mathbf u$, where
\begin{align}
 \mathbf u^{\mathsf T}H\mathbf u
 ={}&2K\sum_i u_i^2
 +J_1\sum_i\sigma_i^{(1)}(u_{i+1}-u_i)^2
 \nonumber\\
 &+J_2\sum_i\sigma_i^{(2)}(u_{i+2}-u_i)^2,
 \label{eq:nnn_quadratic}
\end{align}
and
\begin{equation}
 \sigma_i^{(d)}
 =\cos(\varphi_{i+d}^{(0)}-\varphi_i^{(0)})=\cos\Delta_{i+d}\in\{-1,+1\}\, .
\end{equation}
All bonds have $\sigma_i^{(d)}=+1$ except the nearest-neighbor bond $(0,1)$ and the two next-nearest-neighbor bonds $(-1,1)$ and $(0,2)$, for which $\sigma_i^{(d)}=-1$. In the uniform ferromagnetic regions the nonzero Hessian elements are
\begin{eqnarray}
 H_{ii}&=&2K+2J_1+2J_2,\nonumber\\
 H_{i,i\pm1}&=&-J_1,\qquad
 H_{i,i\pm2}=-J_2.
 \label{eq:nnn_uniform_hessian}
\end{eqnarray}
Thus the 1D ferromagnetic film matrix is a pentadiagonal Toeplitz matrix, but the full domain-wall Hessian is not a Toeplitz matrix: the three reversed bonds form a localized finite-rank perturbation.

\subsection{Uniform spectrum and Chebyshev roots}
\label{app:nnn_bulk}

The eigenvalue equation of the uniform chain is
\begin{align}
 \lambda u_i={}&(2K+2J_1+2J_2)u_i
 -J_1(u_{i-1}+u_{i+1})\nonumber\\
 &-J_2(u_{i-2}+u_{i+2}).
 \label{eq:nnn_bulk_rec}
\end{align}
For a plane wave $u_i=e^{iqia}$, its dispersion is
\begin{equation}
 \lambda(q)=2K+2J_1[1-\cos(qa)]
                 +2J_2[1-\cos(2qa)].
 \label{eq:nnn_dispersion}
\end{equation}
The ferromagnetic (FM)  exchange ground state, corresponding to $q=0$, persists for $J_1>0$ and $J_2\geq-J_1/4$. Indeed, the exchange energy per spin of a planar spiral relative to the FM state is
\begin{align}
 \Delta\varepsilon(q)
 &=\frac{\lambda(q)-\lambda(0)}{2}\nonumber\\
 &=J_1[1-\cos(qa)]+J_2[1-\cos(2qa)]\nonumber\\
 &=2\sin^2\!\frac{qa}{2}\,
   \left[J_1+4J_2\cos^2\!\frac{qa}{2}\right]\, ,
 \label{eq:nnn_ferro_condition}
\end{align}
which is nonnegative for every $q$ precisely when $J_2\geq-J_1/4$ (for $J_1>0$). For long wavelengths,
\begin{equation}
 \Delta\varepsilon(q)=\frac{a^2}{2}(J_1+4J_2)q^2
 +O(q^4).
 \label{eq:nnn_longwave}
\end{equation}
Comparing this expansion with the one-dimensional micromagnetic exchange energy $E_{\mathrm{ex}}=\Ab\int dx\,(\partial_x\mathbf m)^2$ gives the exchange stiffness
\begin{equation}
 \Ab=\frac{a}{2}(J_1+4J_2)
 \label{eq:nnn_stiffness}
\end{equation}
where $a$ is the nearest-neighbor spacing and the spins have unit length. At $J_2=-J_1/4$, $\Ab$ vanishes, although the FM state remains an exchange ground state. The leading exchange-energy dispersion is then $\Delta\varepsilon(q)=J_1a^4q^4/8+O(q^6)$. Throughout this ferromagnetic regime, the bottom of the uniform Hessian spectrum is $2K>0$. Any instability of the EAS must therefore arise from a localized mode below this band.

For a localized eigenvector, put $u_i\propto r^i$ in the uniform recurrence. Introducing
\begin{equation}
 t=2K-\lambda,\qquad x=\frac12(r+r^{-1}),
\end{equation}
gives
\begin{equation}
 t=2J_1(x-1)+4J_2(x^2-1),
 \label{eq:nnn_x_equation}
\end{equation}
or
\begin{equation}
 4J_2x^2+2J_1x-(2J_1+4J_2+t)=0.
 \label{eq:nnn_x_quadratic}
\end{equation}
For $J_2\neq0$, the two solutions are
\begin{equation}
 x_\pm(t)=
 \frac{-J_1\pm\sqrt{(J_1+4J_2)^2+4J_2t}}{4J_2}.
 \label{eq:nnn_x_roots}
\end{equation}
Each $x_\pm$ determines a reciprocal pair $r+r^{-1}=2x_\pm$. 

For $J_2>0$, for a bound mode below the bulk band, $t>0$, $x_+>1$ and $x_-<-1$. The corresponding roots with $|r_\pm|<1$ are
\begin{eqnarray}
 r_+(t)&=x_+-\sqrt{x_+^2-1},\nonumber\\
 r_-(t)&=x_-+\sqrt{x_-^2-1}.
 \label{eq:nnn_r_roots_explicit}
\end{eqnarray}
Including antiferromagnetic $-J_1/4\le J_2$,  we define $r_\pm$ as the unique member of the corresponding pair satisfying $|r_\pm|<1$:
\begin{equation}
 r_\pm=x_\pm-\sqrt{x_\pm-1}\sqrt{x_\pm+1},
 \qquad |r_\pm|<1,
 \label{eq:nnn_r_roots}
\end{equation}
where the square-root product is analytically continued with its branch chosen so that the displayed modulus condition holds. Equivalently, the latter condition defines the roots without any branch convention. For $J_2>0$, $x_+>1$ and $x_-<-1$, so $r_+>0$ and $r_-<0$. For $-J_1/4\leq J_2<0$, the discriminant
\begin{equation}
 \Delta(t)=(J_1+4J_2)^2+4J_2t
 \label{eq:nnn_discriminant}
\end{equation}
may be positive or negative. When $\Delta>0$, both $x_\pm>1$ and the two decaying roots are real and positive. When $\Delta<0$, the roots form a complex conjugate pair $r_-=r_+^*$, with $|r_\pm|<1$. The corresponding real-space tail then has the form $\rho^n[A\cos(n\theta)+B\sin(n\theta)]$, where $r_\pm=\rho e^{\pm i\theta}$. Thus antiferromagnetic $J_2$ can produce exponentially damped oscillations without invalidating the localized-mode construction.

The quartic characteristic equation is reciprocal, and therefore factorizes into two second-order recurrences with parameters $x_\pm$. Each has the Chebyshev form $v_{n+1}=2x_\pm v_n-v_{n-1}$, whose solutions are expressible through the Chebyshev polynomials $U_n(x_\pm)$ of the second kind, as discussed for the tridiagonal problem in the following Appendix~\ref{app:Tridiagonal_Matrix}. For the infinite chain, the decaying roots in Eq.~\eqref{eq:nnn_r_roots} provide the more economical representation.

\subsection{Localized zero mode and critical anisotropy}
\label{app:nnn_secular}

As we discuss in detail in Appendix~\ref{app:EigV_Hessian}, the Hessian is invariant under mirror reflection $i\mapsto 1-i$. We now establish that its lowest eigenmode is antisymmetric
throughout the ferromagnetic exchange regime: 
\begin{equation}
 u_{1+n}=-u_{-n},\qquad n=0,1,\ldots .
 \label{eq:nnn_antisym}
\end{equation}
For $J_2\geq0$, changing the sign of the angular deviations on one side of the wall makes every nonzero off-diagonal Hessian element nonpositive. Since $J_1>0$, the transformed matrix is irreducible; its lowest bound-state eigenvector can be chosen strictly positive. The original lowest eigenvector is therefore reflection-antisymmetric, \textit{i.e.}\ of odd parity.

For $-J_1/4\leq J_2<0$, a different argument is needed. Let $L_0$ denote the exchange Hessian of the homogeneous ferromagnet (\eqref{eq:nnn_energy} with $K=0$), and let $\mathbf d_{ij}=\mathbf e_j-\mathbf e_i$. The exchange Hessian of the EAS is exactly
\begin{align}
 L_{\mathrm{EAS}}={}L_0&
 -2J_1\mathbf d_{01}\mathbf d_{01}^{\mathsf T}
 \nonumber\\
 &-2J_2\bigl(
 \mathbf d_{-1,1}\mathbf d_{-1,1}^{\mathsf T}
 +\mathbf d_{0,2}\mathbf d_{0,2}^{\mathsf T}\bigr).
 \label{eq:nnn_rankone}
\end{align}
In the ferromagnetic regime, $L_0$ is positive semidefinite, as follows directly from Eq.~\eqref{eq:nnn_dispersion} with $K=0$. For $J_2<0$, the last two terms of Eq.~\eqref{eq:nnn_rankone} are positive semidefinite. Hence $L_{\mathrm{EAS}}$ is a positive-semidefinite operator minus a single rank-one negative term, and has at most one negative eigenvalue. For a reflection-symmetric vector $u_{1-i}=u_i$, the nearest-neighbor defect term vanishes because $u_1-u_0=0$; the remaining quadratic form is nonnegative. Any negative eigenmode must therefore  be antisymmetric. Such a negative eigenvalue indeed exists: inserting the antisymmetric trial profile $u_{-n}\propto3^{-n}$, $u_{1+n}=-u_{-n}$ gives
\begin{equation}
 \frac{\mathbf u^{\mathsf T}L_{\mathrm{EAS}}\mathbf u}
 {\mathbf u^{\mathsf T}\mathbf u}
 =-\frac43J_1-\frac{64}{81}J_2<0
 \quad\text{for }-\frac14J_1\leq J_2<0.
 \label{eq:nnn_trial}
\end{equation}
Thus the negative eigenvalue is unique and belongs to the antisymmetric sector. This establishes the mode selection analytically for the entire regime $J_2\geq-J_1/4$, $J_1>0$.

On the left half-chain, the general decaying solution is
\begin{equation}
 u_{-n}=A r_+^n+B r_-^n,\qquad n\geq0.
 \label{eq:nnn_mode}
\end{equation}
Only the equations at sites $0$ and $-1$ differ from the uniform recurrence. Using Eq.~\eqref{eq:nnn_antisym}, they become
\begin{align}
 t u_0
 &-J_1(u_0+u_{-1})
 -J_2(u_{-1}+u_{-2})=0,
 \label{eq:nnn_boundary0}\\
 t u_{-1}
 &+J_1(2u_{-1}-u_0-u_{-2})
 -J_2(u_0+u_{-3})=0.
 \label{eq:nnn_boundary1}
\end{align}
Define
\begin{align}
 F_0(r;t)&=t-J_1(1+r)-J_2(r+r^2),\nonumber\\
 F_1(r;t)&=tr+J_1(2r-1-r^2)-J_2(1+r^3).
 \label{eq:nnn_F}
\end{align}
Substitution of Eq.~\eqref{eq:nnn_mode} into the two
boundary equations yields a homogeneous $2\times2$
system for $(A,B)$. Its determinant must vanish:
\begin{equation}
 F_0(r_+;t)F_1(r_-;t)
 -F_0(r_-;t)F_1(r_+;t)=0.
 \label{eq:nnn_secular}
\end{equation}
Here $r_\pm=r_\pm(t)$ are given by Eqs.~\eqref{eq:nnn_x_roots} and \eqref{eq:nnn_r_roots}.
This is an exact scalar secular equation for the lowest localized eigenvalue in the thermodynamic limit, for either sign of $J_2$. When the discriminant
$\Delta(t)$ vanishes, the roots coalesce,
$r_+=r_-=r$, and the independent solutions become
$r^n$ and $nr^n$. In that case the determinant
in Eq.~\eqref{eq:nnn_secular} vanishes identically
and must be replaced by its confluent form,
\begin{equation}
 F_0(r;t)\,\partial_rF_1(r;t)
 -F_1(r;t)\,\partial_rF_0(r;t)=0.
 \label{eq:nnn_confluent}
\end{equation}
This follows by taking the limit of the original determinant divided by $r_--r_+$.
At $J_2=0$ the characteristic equation drops to second order and is treated separately below. For complex-conjugate roots, Eq.~\eqref{eq:nnn_secular} is purely imaginary; dividing by $r_--r_+$ produces an equivalent real secular equation. Together with the expressions for $r_\pm$, this gives an exact implicit analytical equation for $K_c(J_1,J_2)$ in the thermodynamic limit. It reduces the infinite-dimensional stability problem to a scalar equation involving radicals.

An important simplification is that the anisotropy enters the Hessian only through the uniform shift $2K$. Consequently, the  physically relevant positive solution $t=t_\mathrm{c}$ of Eq.~\eqref{eq:nnn_secular} depends only on $J_1$ and $J_2$, and the lowest EAS Hessian eigenvalue is
\begin{equation}
 \lambda_{\mathrm{EAS}}=2K-t_\mathrm{c}(J_1,J_2).
 \label{eq:nnn_lambda}
\end{equation}
The critical anisotropy and exchange-to-anisotropy ratio are therefore
\begin{equation}
 K_\mathrm{c}(J_1,J_2)=\frac{1}{2}t_\mathrm{c}(J_1,J_2),\qquad
 \alpha_\mathrm{c}\!\left(\frac{J_2}{J_1}\right)
 =\frac{J_1}{K_\mathrm{c}(J_1,J_2)}.
 \label{eq:nnn_critical}
\end{equation}
The EAS is locally stable for $K>K_\mathrm{c}$, marginal at
$K=K_\mathrm{c}$, and unstable for $K<K_\mathrm{c}$.
The expression is implicit but exact and involves
only two decaying roots and a $2\times2$ determinant,
rather than the determinant of an infinite pentadiagonal
matrix~\cite{Elouafi:2018,Andelic:2021}.

\subsection{Nearest-neighbor limit and weak next-nearest-neighbor exchange}
\label{app:nnn_limits}

For $J_2=0$, only one decaying root survives.
The first boundary equation reduces to
$t=J_1(1+r)$, while the bulk recurrence gives
$t=J_1(r+r^{-1}-2)$.
Together these yield
\begin{equation}
 r=\frac13,\qquad
 t_\mathrm{c}=\frac43J_1,\qquad
 K_\mathrm{c}=\frac23J_1,\qquad
 \alpha_\mathrm{c}=\frac32,
 \label{eq:nnn_nn_limit}
\end{equation}
recovering Appendix~A.

The first-order correction for small $|J_2|/J_1$ can be obtained directly from the unperturbed normalized zero-mode profile, proportional to $(1/3)^n$ on each side of the central bond. The next-nearest-neighbor contribution to its Rayleigh quotient is
\begin{equation}
 \left.\frac{\partial\lambda_{\mathrm{EAS}}}
 {\partial J_2}\right|_{J_2=0}
 =
 \frac{\displaystyle
 \sum_{n=0}^{\infty}(r^n-r^{n+2})^2
 -(1+r)^2}
 {\displaystyle\sum_{n=0}^{\infty}r^{2n}}
 \bigg|_{r=1/3}
 =-\frac{64}{81}.
 \label{eq:nnn_perturb}
\end{equation}
The negative term is the combined contribution of the two reversed next-nearest-neighbor bonds crossing the wall, after reduction to one half-chain. It follows that
\begin{equation}
 K_\mathrm{c}=\frac23J_1+\frac{32}{81}J_2
 +O\!\left(\frac{J_2^2}{J_1}\right),
 \label{eq:nnn_K_expansion}
\end{equation}
or, equivalently,
\begin{equation}
 \alpha_\mathrm{c}=\frac32\left[
 1-\frac{16}{27}\frac{J_2}{J_1}
 +O\!\left(\frac{J_2^2}{J_1^2}\right)
 \right].
 \label{eq:nnn_alpha_expansion}
\end{equation}
For $J_1\to0^+$ at fixed $J_2$, the even and odd
sublattices decouple into nearest-neighbor chains
with exchange $J_2$. Each has a reversed bond at the
EAS, so $\alpha_\mathrm{c}\to 3J_2/(2K)$.

\begin{table}
\centering
\caption{Numerical solution of the analytical
secular equation~\eqref{eq:nnn_secular} including its confluent limit \eqref{eq:nnn_confluent}, to determine the critical anisotropy $K_\mathrm{c}$ in units of $J_1$ and $\alpha_\mathrm{c}=(J_1/K)_\mathrm{c}$. \label{tab:app-numerical-values} required to stabilize the EAS of the domain wall.}
\begin{tabular}{c@{\hskip 5mm} c@{\hskip 5mm}c}
\hline\hline
$J_2/J_1$ & $K_\mathrm{c}/J_1$ & $(J_1/K)_\mathrm{c}$\\
\hline
$-0.25$ & $0.589540$ & $1.696239$\\
$-0.20$ & $0.600709$ & $1.664699$\\
$-0.10$ & $0.630080$ & $1.587101$\\
$0$   & $0.666667$ & $1.500000$\\
\phantom{$-$}$0.10$ & $0.708568$ & $1.411297$\\
\phantom{$-$}$0.50$ & $0.907978$ & $1.101348$\\
\phantom{$-$}$1.00$   & $1.192704$ & $0.838431$\\
\phantom{$-$}$2.00$   & $1.806331$ & $0.553608$\\
\hline\hline
\end{tabular}
\end{table}
For illustration, \eqref{eq:nnn_K_expansion}, \eqref{eq:nnn_alpha_expansion} and Table~\ref{tab:app-numerical-values} show that a positive next-nearest-neighbor exchange increases the critical anisotropy required to stabilize the atomically sharp easy-axis wall, whereas competing antiferromagnetic $J_2$ decreases it, as long as the ferromagnetic state remains stable. For a fixed $J_1>0$, the critical ratio $\alpha=J_1/K$, therefore, decreases (increases) with increasing (decreasing) $J_2$. The weak-$J_2$ expansion \eqref{eq:nnn_K_expansion} applies on both sides of $J_2=0$. The instability is a localized, reflection-antisymmetric mode throughout the entire ferromagnetic regime, although its tail can be oscillatory for negative $J_2$.

The positive next-nearest-neighbor exchange lowers the critical ratio $J_1/K$ by increasing the anisotropy required to stabilize the collinear EAS. Thus the critical value $\alpha_\mathrm{c}=3/2$ is specific to the nearest-neighbor model, although the instability remains associated with a localized Hessian mode. The generalized critical ratio is not simply obtained by intuitively replacing $J$ with $J_1+4J_2$, as a continuum exchange-stiffness argument might suggest. The exact result depends on the localized bonds crossing the wall.

\section{Tridiagonal matrix and  Chebyshev polynomials}
\label{app:Tridiagonal_Matrix}

We consider the uniform tridiagonal matrix, 
\begin{equation}
 \matr{T}_n(2\lambda)=
 \begin{pmatrix}
 2\lambda&-1&&\\
 -1&2\lambda&-1&\\
 &\ddots&\ddots&\ddots\\
 &&-1&2\lambda
 \end{pmatrix},
 \label{eq:app_T}
\end{equation}
with $n=N-2$ degrees of freedom, spins at site $i=1$ and $i=N$ is fixed to out-of-plane direction  and
\begin{equation}
 \lambda=1+\frac{1}{\alpha}.
 \label{eq:app_lambda}
\end{equation}

\subsection{Determinant of tridiagonal matrix}
Defining the determinant of $\matr{T}_n$
\begin{equation}
 D_n(\lambda):=\det\matr{T}_n(2\lambda),
\end{equation}
and expanding according to the Laplace rule along the first row yields
\begin{equation}
 D_n(\lambda) =2\lambda D_{n-1}(\lambda)  -D_{n-2}(\lambda),
 \label{eq:app_recurrence}
\end{equation}
with $D_0=1$ and $D_1=2\lambda$. This is precisely the recurrence relation of the Chebyshev polynomials of the second kind,
$U_n(\lambda)$, with identical initial conditions. Consequently,
\begin{equation}
 D_n(\lambda):=U_n(\lambda)\, .
\label{eq:app_det_Chebyshev}
\end{equation}

\subsection{Eigenvalues of tridiagonal matrix}
The eigenvalues $\epsilon_m$ of the tridiagonal matrix $\matr{T}_n$ in \eqref{eq:app_T}
\begin{equation}
(\matr{T}_n- \epsilon_m \matr{1}_n)\mathbf{v}^{(m)}=0
\label{eq:app_T-Eigenvalue-problem}
\end{equation}
are the zeros of the determinant and the zeros of the Chebychev polynomials 
\begin{equation}
 D_n(\lambda- \epsilon_m/2)=U_n(\lambda- \epsilon_m/2)=0\, .
\label{eq:app_det_Chebyshev=0}
\end{equation}
Employing the well-known explicit representation of $U_n(x)$ for $|x|\le 1$
\begin{equation}
U_n(x) = \frac{\sin{(n+1)\tau}}{\sin{\tau}}\quad \text{with} \quad x=\cos{\tau}\quad (0\le\tau < \pi)\, ,
\label{eq:Cheby2nd_representation}
\end{equation}
the determinant is zero for all  $\tau_m= \frac{m}{n+1}\pi$ with $m=1,\dots , n$ and the eigenvalues are determined from  $x= \lambda -\epsilon_m/2 = \cos\tau_m$,
\begin{equation}
\epsilon_m=2\lambda-2\cos\tau_m\quad\text{with}\quad\tau_m= \frac{m}{n+1}\pi\, .
\label{eq:app_eigenvalue_epsilon}
\end{equation}
Thus, for $\lambda>0$, the eigenvalue spectrum is bound between $2\lambda-2 \le \epsilon_m \le 2\lambda+2$.
\subsection{Eigenvectors of tridiagonal matrix}
We seek eigenvectors $\mathbf{v}^{m}$ that satisfy \eqref{eq:app_T-Eigenvalue-problem}. For the interior sites, \eqref{eq:app_T-Eigenvalue-problem} gives the difference equation
\begin{equation}
- v_{i-1}^{(m)} +(2\lambda-\epsilon_m) v_{i}^{(m)}  -v_{i+1}^{(m)}=0\, .
\label{eq:app_eigenvector_recursion}
\end{equation}
Identifing $\lambda-\epsilon_m/2=x_m$,  it is exactly the recursion relation of the Chebychev polynomials \eqref{eq:app_recurrence} and with $v_0^{(m)}=0$ for all eigenvectors $(m)$, using \eqref{eq:Cheby2nd_representation} and \eqref{eq:app_eigenvalue_epsilon} the eigenvectors can be expressed by the Chebychev polynomials
\begin{eqnarray}
v_{i+1}^{(m)}&=&v_{1}U_i(x_m)=v_{1}U_i(\lambda-\epsilon_m/2)=v_{1}U_i(\cos\tau_m) \nonumber \\
&=& v_{1}\frac{\sin{(i+1)\tau_m}}{\sin{\tau_m}}
\label{eq:app_eigenvector_chain}
\end{eqnarray}
We conclude, the eigenvectors  have a particularly simple form: they are discrete sine waves, and their components can also be expressed in terms of Chebyshev polynomials of the second kind. With the proper normalization we can write them finally as 
\begin{equation}
v_{i}^{(m)}= \sqrt{\frac{2}{n+1}}\sin{\left(\frac{im}{n+1}\pi\right)}\,\quad i,m=1,\cdots,n\, .
\end{equation}
Thus, the same Chebyshev polynomials determine both the eigenvalues and the eigenvectors. An important observation is that the eigenvectors do not depend on $\lambda$. Changing $\lambda$ shifts all eigenvalues by the same amount without changing the eigenvectors. The lowest mode, $m=1$, has no internal nodes, successive modes, $m>1$ have increasing numbers of nodes.

\section{Symmetry and localization of the Hessian modes without and with DMI}
\label{app:EigV_Hessian}

\subsection{Hessian and eigenvalue decomposition}

Here we resolve the two lowest Hessian modes by their mirror parity. Fig.~\ref{fig:appendix-eigenvalues} presents the  mode-resolved eigenvalues of Hessian matrix as the function of $\alpha$. Fig.~\ref{fig:appendix-modes-d0} and  Fig.~\ref{fig:appendix-modes-d02} present real space distributions of relaxed chains and the two lowest mode-resolved Hessian eigenvectors when $D/J=0$ and 0.2, respectively. The distributions of eigenvectors reveal that as the wall narrows, A mode becomes localized and  concentrated on the central bond. However, S mode becomes flat and extended in the EAS without DMI.  To illustrate their evolution as the wall narrows, we first define the bond stiffness for a relaxed planar configuration
\begin{equation}
\kappa_i
=\frac{\partial^2}{\partial\Delta_i^2}
\left[-J\cos\Delta_i+D_y\sin\Delta_i\right]
=J\cos\Delta_i-D_y\sin\Delta_i.
\label{eq:bond-stiffness}
\end{equation}
The nonzero Hessian elements are
\begin{equation}
H_{ii}=\kappa_{i-1}+\kappa_i-2K\cos 2\varphi_i,
\qquad
H_{i,i+1}=-\kappa_i.
\label{eq:relaxed-hessian-elements}
\end{equation}
For a normalized eigenvector $\sum_i u_i^2=1$, its eigenvalue is the Rayleigh quotient
\begin{align}
\lambda
&=\mathbf u^{\mathsf T}\matr H\mathbf u \nonumber\\
&=\sum_i\kappa_i(u_{i+1}-u_i)^2
-2K\sum_i\cos(2\varphi_i)u_i^2. 
\label{eq:hessian-rayleigh}
\end{align}
Using $\cos(2\varphi_i)=2S_{x,i}^2-1$ and the normalization of $\mathbf u$ gives the following decomposition
\begin{equation}
\lambda=2K+E_{\rm well}+E_{\rm bond},
\label{eq:hessian-decomposition}
\end{equation}
with
\begin{equation}
E_{\rm well}=-4K\sum_iS_{x,i}^2u_i^2,
\qquad
E_{\rm bond}=\sum_i\kappa_i(u_{i+1}-u_i)^2.
\end{equation}
Thus, $2K$ is the lower edge of the uniform-domain spectrum. The term $E_{\rm well}$ binds a mode where the relaxed wall has a transverse component $S_x$, whereas $E_{\rm bond}$ becomes attractive on a sharp wall for which $\kappa_i<0$.

\subsection{Symmetry classification and localization of the Hessian modes}

\begin{figure}[tp]
\includegraphics[width=0.48\textwidth]{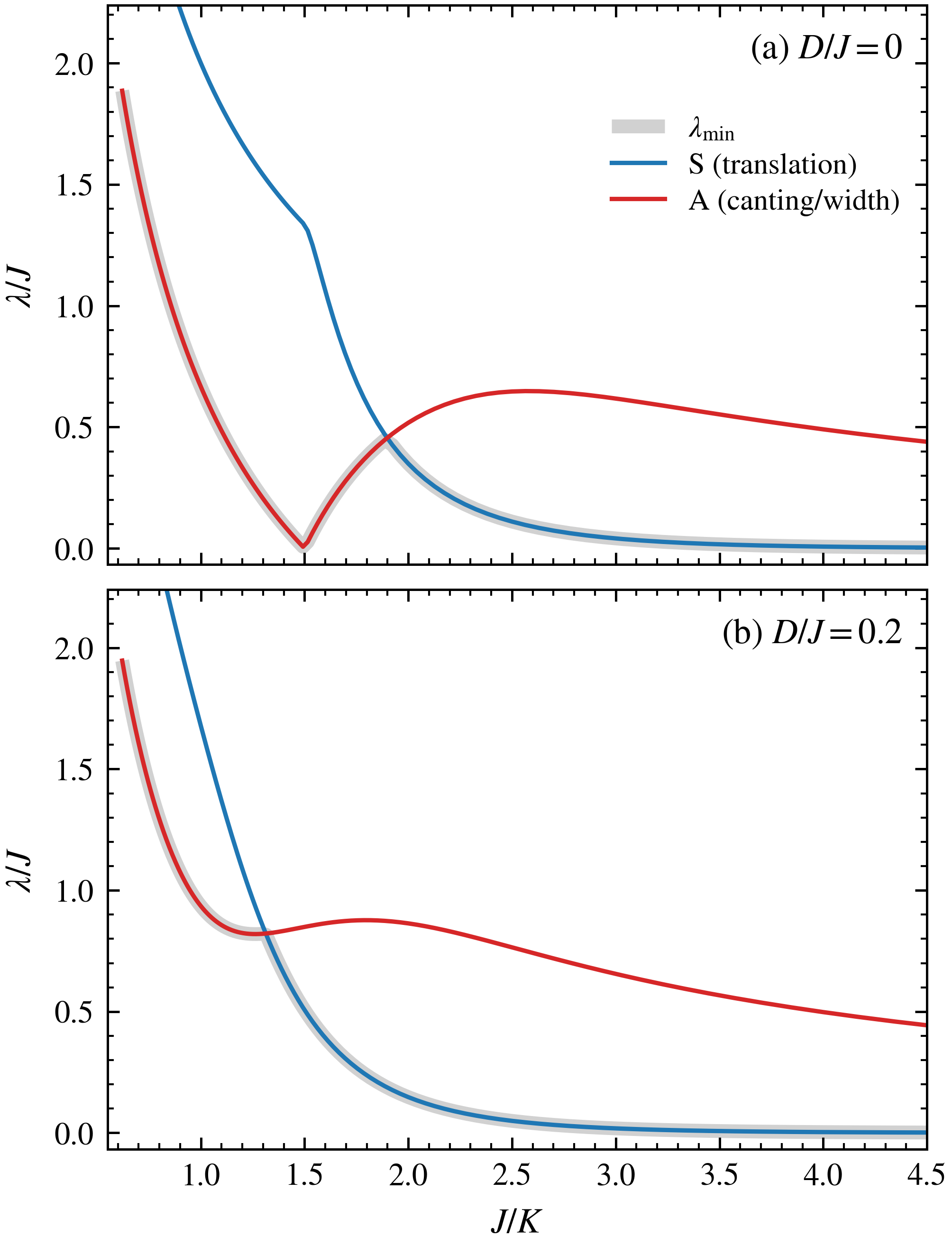}
\caption{Lowest symmetric (S) and antisymmetric (A) Hessian eigenvalues as functions of $J/K$ for (a) $D_y=0$ and (b) $D_y=0.2J$. The thick gray curve is $\lambda_{\min}=\min(\lambda_S,\lambda_A)$. Its kinks occur where the symmetry-resolved branches cross. The nearly flat interval  around $\lambda/J \approx 0.8$ in panel (b) follows the shallow minimum of the A branch.}
\label{fig:appendix-eigenvalues}
\end{figure}

Because the Hessian is mirror symmetric about the central bond, its eigenvectors can be chosen as symmetric, $u_i=u_{N+1-i}$, or antisymmetric, $u_i=-u_{N+1-i}$. We denote the lowest eigenvalue in these two subspaces by $\lambda_S$ and $\lambda_A$, respectively. The plotted lowest eigenvalue is simply $\lambda_{\min}=\min(\lambda_S,\lambda_A)$. Figure~\ref{fig:appendix-eigenvalues} shows that every kink in $\lambda_{\min}$ is an S--A crossing. 

In the uniform tails of either domain, a bound eigenvector has the form $u_n\propto r^{|n|}$ with $0<r<1$. Substitution into the bulk recurrence gives
\begin{equation}
2K-\lambda=J\left(r+r^{-1}-2\right)
=2J\left(\cosh\gamma-1\right),
\qquad r=e^{-\gamma}.
\label{eq:hessian-localization-length}
\end{equation}
Thus, binding is determined relative to the bulk edge $2K$: a localized mode below this edge, is bound,  has a binding energy $E_{\mathrm{bind}}=2K-\lambda>0$ and a localization length $\xi_{\mathrm{loc}}=1/\gamma$, measured in lattice $\lambda<2K$, is bound, and the localization length is $\xi_{\rm loc}=1/\gamma$ expressed in units of the lattice spacing. A more negative value of $\lambda-2K$ implies a shorter asymptotic tail and a sharper wall. However, the binding energy alone does not reveal the microscopic origin of localization or distinguish between symmetric and antisymmetric modes.

\subsection{Evolution of the Hessian modes and domain-wall profiles}


\begin{figure}[tbp]
\centering
\includegraphics[width=0.96\columnwidth]{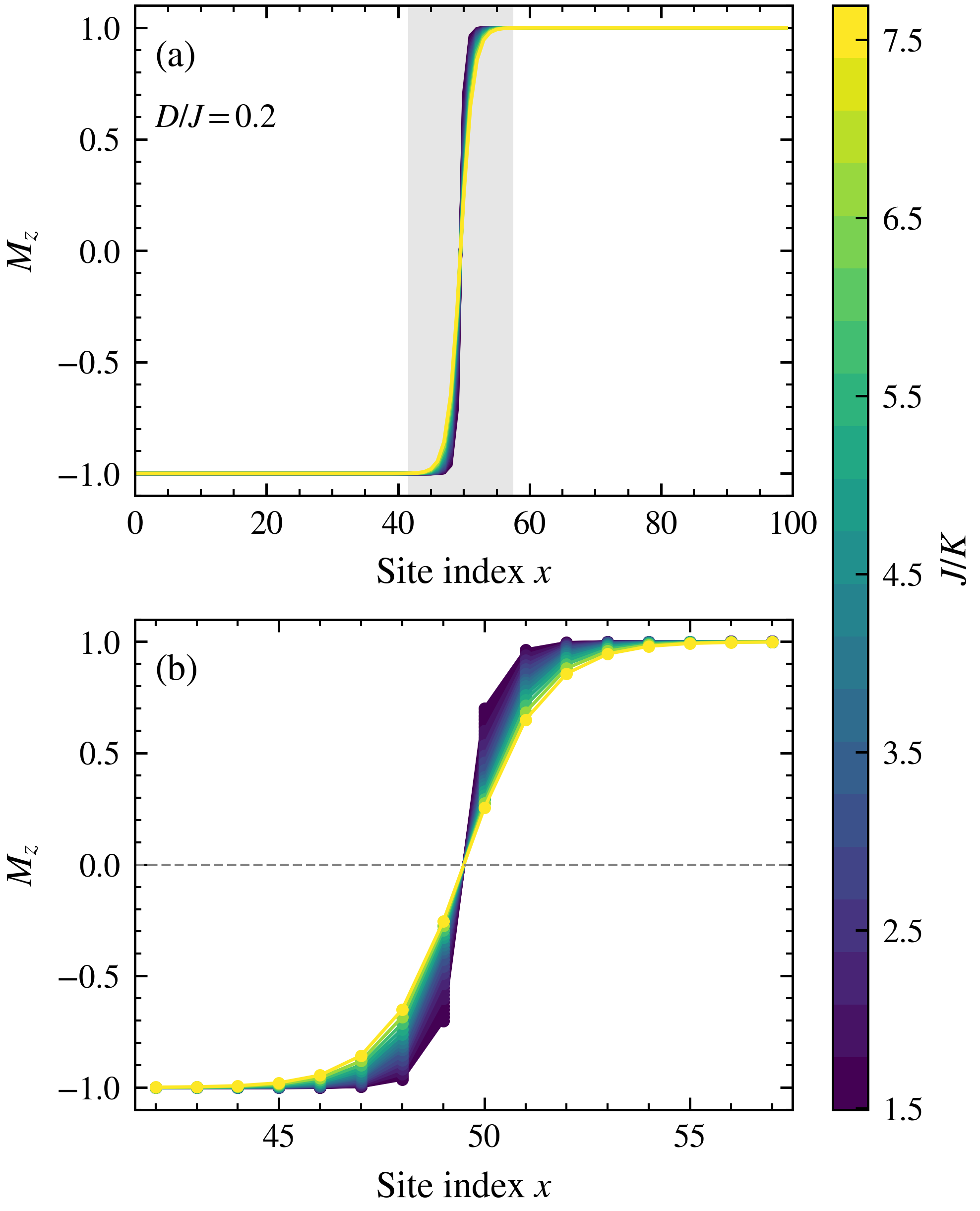}
\caption{Out-of-plane spin component of relaxed chains for $D_y=0.2J$ over $J/K$. Panel (a) shows the full chain and panel (b) enlarges the wall center. The profiles evolve continuously through the region where the lowest Hessian branch changes mirror parity, confirming that the kink is a level crossing rather than switching between distinct relaxed walls.}
\label{fig:appendix-profiles}
\end{figure}
\begin{figure*}[tp]
\centering
\includegraphics[width=0.68\textwidth]{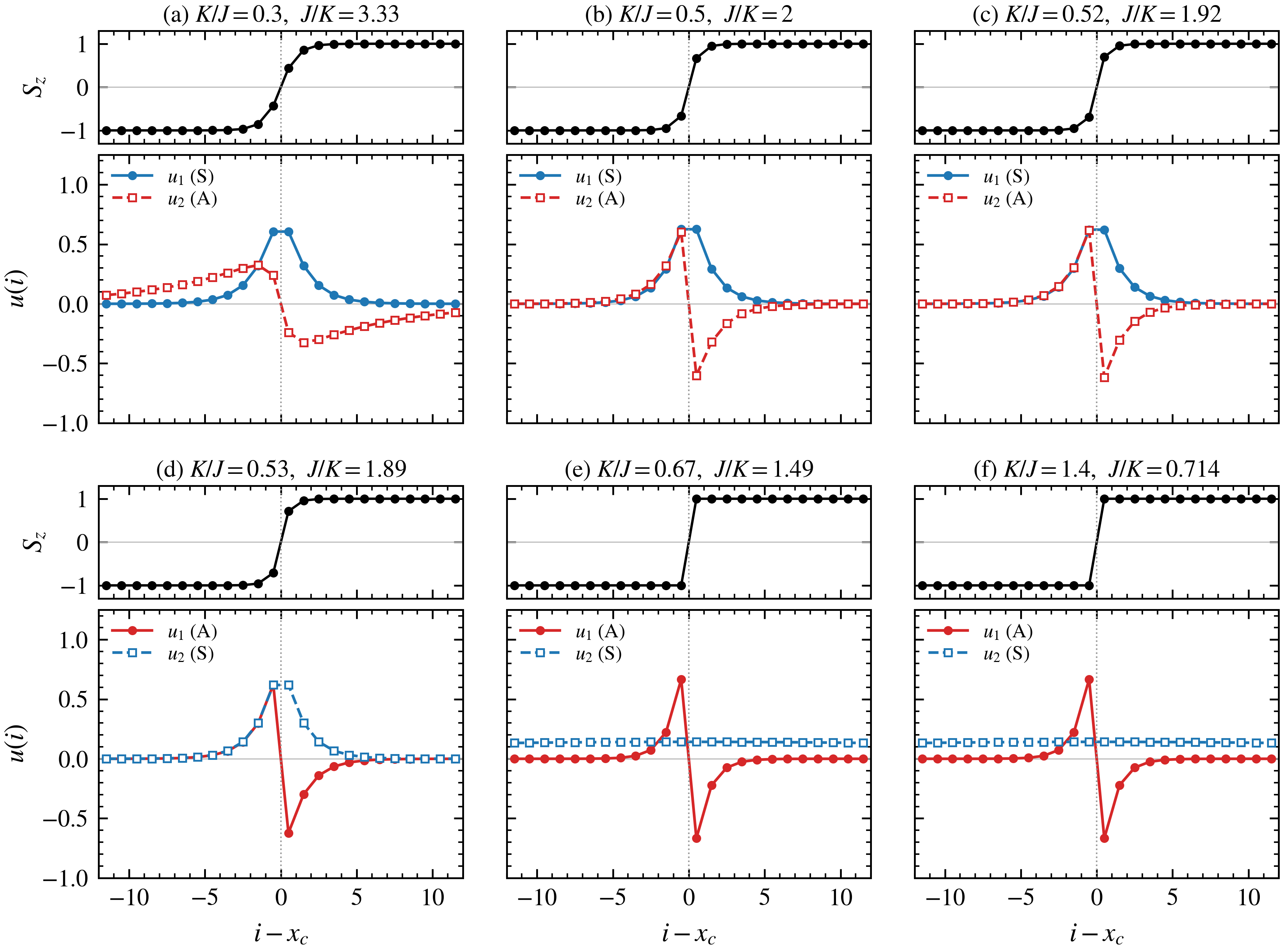}
\caption{Relaxed profiles (upper panels) and the two lowest mode-resolved Hessian eigenvectors (lower panels) for $D_y=0$ at representative anisotropies. As the wall becomes atomically sharp, the S mode spreads through the domains, whereas the A mode becomes concentrated on the central bond.}
\label{fig:appendix-modes-d0}
\end{figure*}
\begin{figure*}[tp]
\centering
\includegraphics[width=0.68\textwidth]{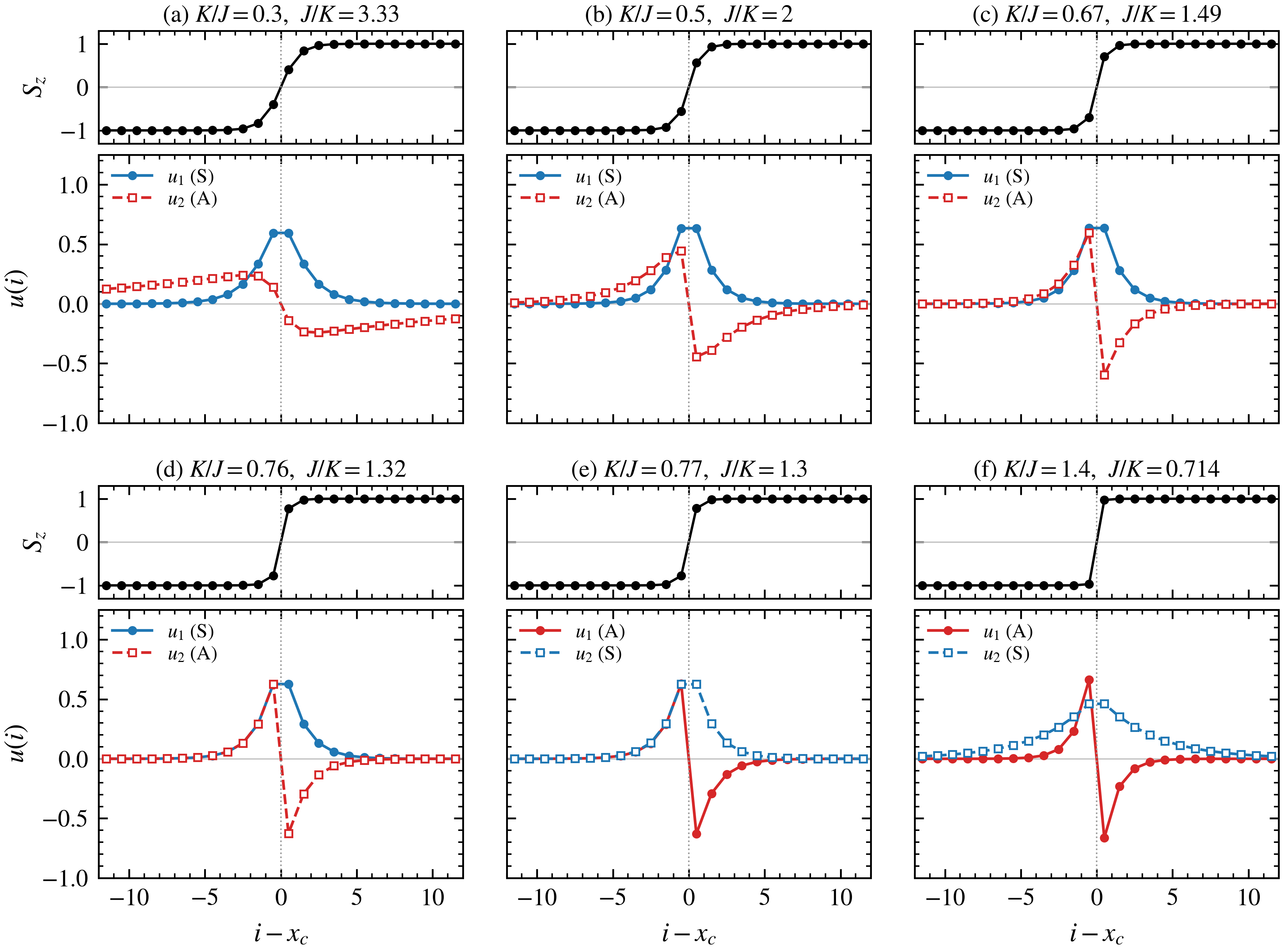}
\caption{Same as Fig.~\ref{fig:appendix-modes-d0}, but for $D_y=0.2J$. Around the S--A crossing the ordering of the two modes changes without a discontinuity of either symmetry-resolved branch. On the sharp-wall side, the A mode remains localized within a few sites of the wall center.}
\label{fig:appendix-modes-d02}
\end{figure*}

The decomposition in \eqref{eq:hessian-decomposition} provides a qualitative physical interpretation of the different localization mechanisms of the symmetric (S) and antisymmetric (A) Hessian modes shown in Fig.~\ref{fig:appendix-modes-d0} and  Fig.~\ref{fig:appendix-modes-d02} in terms of eigenvectors.  The symmetric mode, where $u_1=u_0$, has equal amplitudes on the two sites adjacent to the central bond, so that this bond makes no contribution to $E_{\mathrm{bond}}$. Its localization is therefore primarily governed by the attractive anisotropy-induced potential well $E_{\mathrm{well}}$, which develops wherever the equilibrium magnetization acquires a transverse component. As the domain wall approaches the atomically sharp EAS, this transverse component disappears, the anisotropy well vanishes, and the symmetric mode becomes extended.

The antisymmetric mode, in contrast, changes sign across the central bond, $u_1=-u_0$, and consequently has a finite amplitude difference between its two central sites. It is therefore particularly sensitive to the central bond stiffness $\kappa_i$. As the wall narrows and this stiffness becomes negative, the corresponding contribution to $E_{\mathrm{bond}}$ becomes attractive, favoring a strongly localized mode concentrated around the wall center. The different responses of the two symmetry sectors explain why the lowest Hessian eigenmode can change its mirror parity as the exchange-to-anisotropy ratio is varied.

The competition between the anisotropy-induced potential well and the bond-stiffness contribution also determines the evolution of the symmetry-resolved eigenvalues. In particular, for $D_y=0.2J$, the shallow minimum of the antisymmetric branch produces the nearly flat region of the lowest-eigenvalue curve in Fig.~\ref{fig:appendix-eigenvalues}(b). When the symmetric and antisymmetric branches cross, the identity of the lowest Hessian mode changes, resulting in a kink in $\lambda_{\min}=\min(\lambda_{\mathrm S},\lambda_{\mathrm A})$. Since the two modes belong to different reflection-symmetry sectors, their eigenvalues may cross without hybridization. Importantly, such a crossing concerns the spectrum of fluctuations around the equilibrium domain wall and does not, by itself, indicate a phase transition or a discontinuity in the equilibrium spin configuration. This distinction is confirmed by Fig.~\ref{fig:appendix-profiles}, which demonstrates that the relaxed domain-wall profiles evolve continuously through the parameter range where the lowest Hessian mode changes its mirror parity.

\clearpage

\bibliography{domain_walls_PRB_references}
\end{document}